\documentclass[usenatbib,usegraphicx]{mn2e}
\usepackage{amssymb,color,multirow,charter,eulervm}

\newcommand{\oneby}[1]{\frac{1}{#1}}                      
\newcommand{\e}[1]{\times 10^{#1}}                        

\newcommand{\dd}[2][]{\frac{d #1}{d #2}}                  

\newcommand{\fig}[1]{Fig. \ref{#1}}

\pdfoutput=1

\newcommand{\fignar}[3]{
   \begin{figure}
   \centering
   \includegraphics{#1}
   \caption{#3}
   \label{#2}
   \end{figure}
}

\newcommand{\figwide}[3]{
   \begin{figure*}
   \centering
   \includegraphics{#1}
   \caption{#3}
   \label{#2}
   \end{figure*}
}

\title[Accretion to Magnetized Stars through the Rayleigh-Taylor Instability]
{Accretion to Magnetized Stars through the Rayleigh-Taylor Instability: Global Three-Dimensional Simulations}

\author[A. K. Kulkarni \& M. M. Romanova]
{A. K. Kulkarni\thanks{E-mail: akshay@astro.cornell.edu},
M. M. Romanova\thanks{E-mail: romanova@astro.cornell.edu}\\
Dept. of Astronomy, Cornell University, Ithaca, NY 14853}

\begin{document}
\maketitle
\label{firstpage}

\begin{abstract}
We present results of 3D simulations of MHD instabilities at the accretion disk-magnetosphere boundary. The instability is Rayleigh-Taylor, and develops for a fairly broad range of accretion rates and stellar rotation rates and magnetic fields. It manifests itself in the form of tall, thin tongues of plasma that penetrate the magnetosphere in the equatorial plane. The shape and number of the tongues changes with time on the inner-disk dynamical timescale. In contrast with funnel flows, which deposit matter mainly in the polar region, the tongues deposit matter much closer to the stellar equator. The instability appears for relatively small misalignment angles, $\Theta\lesssim30^\circ$, between the star's rotation and magnetic axes, and is associated with higher accretion rates. The hot spots and light curves during accretion through instability are generally much more chaotic than during stable accretion. The unstable state of accretion has possible implications for quasi-periodic oscillations and intermittent pulsations from accreting systems, as well as planet migration.
\end{abstract}

\begin{keywords}
accretion, accretion discs; instabilities; MHD; stars: oscillations; stars: magnetic fields
\end{keywords}

\section{Introduction}
Magnetospheric accretion occurs in different systems, including classical T Tauri stars (CTTSs), which are the progenitors of solar-type stars (e.g., Hartmann 1998; Bouvier et al. 2007), in magnetized cataclysmic variables, which are accreting white dwarfs (e.g., Warner 1995; Warner \& Woudt 2002), and in millisecond pulsars, which are weakly magnetized accreting neutron stars (e.g., van der Klis 2000). The geometry of the accretion flow around magnetized stars is a problem of long-standing interest. It is an important factor in determining the observed spectral and variability properties of the accreting system. An accretion disk around a magnetized central object is truncated at the distance from the central star where the magnetic energy density becomes comparable to the matter energy density. Beyond that point, there are two ways in which the gas can accrete to the star: (1) Through funnel streams, or magnetospheric accretion (e.g., Ghosh \& Lamb 1978, 1979); (2) Through plasma instabilities at the disk-magnetosphere interface (e.g., Arons \& Lea 1976; Elsner \& Lamb 1977). The geometry of the matter flow in these two regimes is expected to be very different.
In general, the Rayleigh-Taylor (RT), or interchange, instability is expected to develop at the disk-magnetosphere interface because of the high-density disk matter being supported against gravity by the low-density magnetospheric plasma. The Kelvin-Helmholtz (KH) instability is also expected to develop because of the discontinuity in the angular velocity of matter at the boundary. The inner disk matter is expected to rotate at the local Keplerian velocity, while the magnetospheric plasma corotates with the star.

A number of theoretical analyses of such instabilities have been performed. Arons \& Lea (1976a,b) presented a detailed analysis of magnetospheric accretion through the RT instability for a non-rotating star and a spherical accretion geometry. Anzer (1969) and Anzer \& B\"orner (1980, 1983) have analyzed the RT instability for solar prominences and the KH instability at the inner disk edge respectively. Baan (1977, 1979) examined the role of the RT instability in X-ray bursts. Lovelace \& Scott (1981) investigated the interchange instability in a two-dimensional accretion disk threaded by a vertical magnetic field. Spruit \& Taam (1990) investigated the stability of a thin magnetized disk with a rigid rotation profile. They present a solution to one of the problems with the model of Ghosh \& Lamb (1979), namely that the incoming disk squeezes the magnetosphere in the equatorial region, and it is energetically impossible for the inner disk matter to follow the resulting magnetospheric field lines. The solution is that the matter at the inner disk edge can be unstable, causing it to move towards the star across the field lines, until it reaches field lines that are energetically possible to follow. The RT instability in differentially rotating magnetized disks was studied by Spruit, Stehle \& Papaloizou (1995) and Lubow \& Spruit (1995). Li \& Narayan (2004) analyzed RT and KH instabilities at the disk-magnetosphere interface for an infinitely thick disk and a vertical magnetic field.

Theoretical analysis, although useful in understanding the basic features of the instabilities, is limited to the linear regime. To understand the behaviour of the instabilities in the nonlinear regime, numerical simulations are needed. While there have been some numerical studies of such instabilities, they have so far been limited in one way or another, and have mostly focused on the physical characteristics of the instabilities themselves, with relatively less attention being given to their effect on the nature of the accretion flow, and to their dependence on the physical parameters of accreting systems, like the accretion rate and the star's rotation rate and magnetic field. Moreover, realistic accretion geometries can only be obtained from global 3D simulations, which have so far been lacking. 2D simulations have been performed by Wang \& Nepveu (1983), Wang \& Robertson (1984, 1985) and Kaisig, Tajima \& Lovelace (1992). Rast\"atter \& Schindler (1999b) performed 3D simulations in a patch which includes a part of the magnetospheric boundary and the equatorial plane, using semianalytically derived initial conditions (Rast\"atter \& Neukirch 1997; Rast\"atter \& Schindler 1999a). Stone \& Gardiner (2007a,b) performed 3D simulations of the magnetic RT instability in a shearing box for the idealized case of two inviscid, perfectly conducting fluids of constant density separated by a contact discontinuity perpendicular to the effective gravity, with a uniform magnetic field parallel to the interface.

Earlier, two- and three-dimensional simulations have shown accretion through funnel streams (Romanova et al. 2002, 2003, 2004; Kulkarni \& Romanova 2005). In this paper we report on accretion through the Rayleigh-Taylor, or interchange, instability in global 3D MHD simulations, where the simulation region includes the disk and the whole magnetosphere of the star. This paper follows up on an earlier, briefer paper (Romanova, Kulkarni \& Lovelace 2008), giving a more detailed description of the problem, and showing results for a larger range of star and disk parameters.

\section{The Numerical Model}
\label{sec:model}
The model we use is the same as in our earlier 3D MHD simulations (Koldoba et al. 2002; Romanova et al. 2003, 2004). The star has a dipole magnetic field, the axis of which makes an angle $\Theta$ with the star's rotation axis. The rotation axes of the star and the accretion disk are aligned. There is a low-density corona around the star and the disk which also rotates about the same axis. To model stationary accretion, the disk is chosen to initially be in a quasi-equilibrium state, where the gravitational, centrifugal and pressure gradient forces are in balance (Romanova et al. 2002). Simulations were done for both relativistic and non-relativistic stars. General relativistic effects, which are important for neutron stars, are modelled using the Paczy\'nski-Wiita potential $\Phi(r) = GM_*/(r-r_g)$ (Paczy\'nski \& Wiita 1980), where $r_g \equiv 2GM_*/c^2$ is the Schwarzschild radius of the star. This potential reproduces some important features of the Schwarzschild spacetime, such as the positions of the innermost stable and marginally bound circular orbits. Viscosity is modelled using the $\alpha$-model (Shakura \& Sunyaev 1973; Novikov \& Thorne 1973), and controls the accretion rate through the disk. To model accretion, the ideal MHD equations are solved numerically in three dimensions, using a Godunov-type numerical code, written in a ``cubed-sphere'' coordinate system rotating with the star (Koldoba et al. 2002; Romanova et al. 2003). The boundary conditions at the star's surface amount to assuming that the infalling matter passes through the surface of the star. So the dynamics of the matter after it falls on the star is ignored. The inward motion of the accretion disk is found to be stopped by the star's magnetosphere at the Alfv\'en or magnetospheric radius, where the magnetic and matter energy densities become equal. The subsequent evolution depends on the parameters of the model.

\subsection{Reference Values}
\label{sec:refval}
The simulations are done using dimensionless variables. For every physical quantity $q$, the dimensionless value is defined as $q' = q/q_0$, where $q_0$ is the reference value for $q$. Because of the use of dimensionless variables, the results are applicable to a wide range of objects and physical conditions, each with its own set of reference values, provided the magnetospheric radius $r_m$ is not very large compared with the radius of the star $R_*$, $r_m = (4-6)R_*$; $r_m$ is determined by the balance of the magnetospheric and matter pressure, so that the modified plasma parameter at the disk-magnetosphere boundary $\beta=(p+\rho v^2)/(B^2/{8\pi})\approx 1$. To apply our simulation results to a particular situation, we have the freedom to choose three parameters, and all the reference values are calculated from those. We choose the mass, radius and equatorial surface magnetic field of the star as the three independent parameters. Appendix \ref{app:refval} shows how the reference values are determined, and lists the reference values for three classes of central objects --- classical T Tauri stars, white dwarfs and neutron stars.

Subsequently, we drop the primes on the dimensionless variables and show dimensionless values in the figures.

\section{Simulation Results}
We chose the following parameters for our main case: dipole moment $\mu=2$, corresponding to an equatorial surface magnetic field of $B_*=2$, misalignment angle $\Theta=5^\circ$, viscosity parameter $\alpha=0.1$, stellar rotation period $P=3$, initial disk radius $=2$. The corotation radius, which is the radius at which the orbital rotation rate of the disk matter equals the star's rotation rate, is $r_{cor}=2$. The cubed-sphere grid resolution in most of our cases is $N_r\times N^2 = 72\times31^2$ in each of the six blocks of the grid (which is the resolution in all the figures in this paper, unless otherwise noted), and we have some runs at higher resolutions ($100\times41^2, 144\times61^2, 216\times91^2$). The outer radius of our simulation region is $r_{out}\approx12\approx35R_*$ in most runs, although we only show the region near the star in our plots.

\figwide{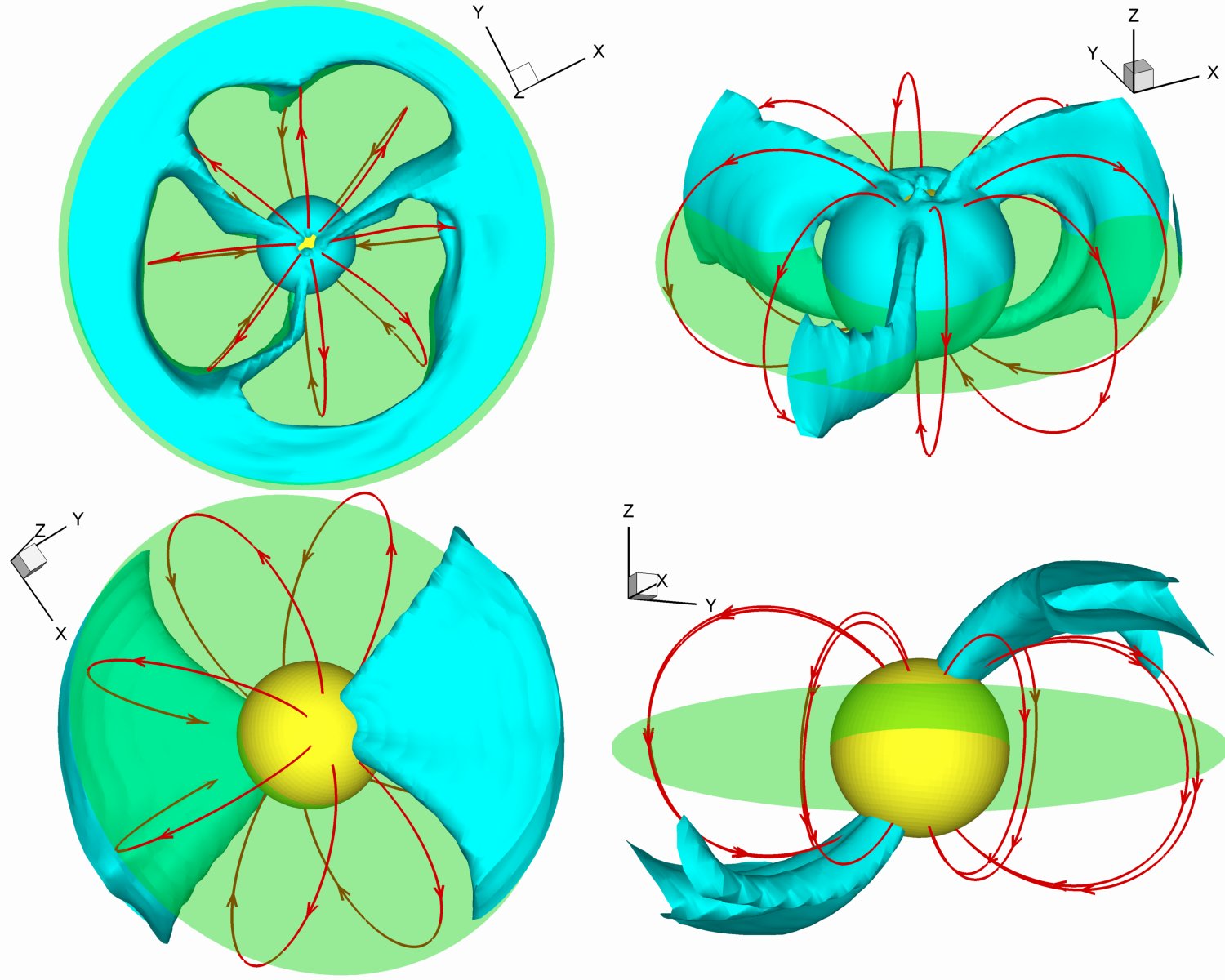}{3d-main}{Geometry of the accretion flow around a star through instabilities (top row), contrasted with a traditional funnel flow (bottom row). A constant density surface is shown. The lines are magnetospheric magnetic field lines. The translucent disc denotes the equatorial plane. The star's rotation axis is in the z-direction (see also http://astro.cornell.edu/us-rus/stereo.htm for animations).}

\figwide{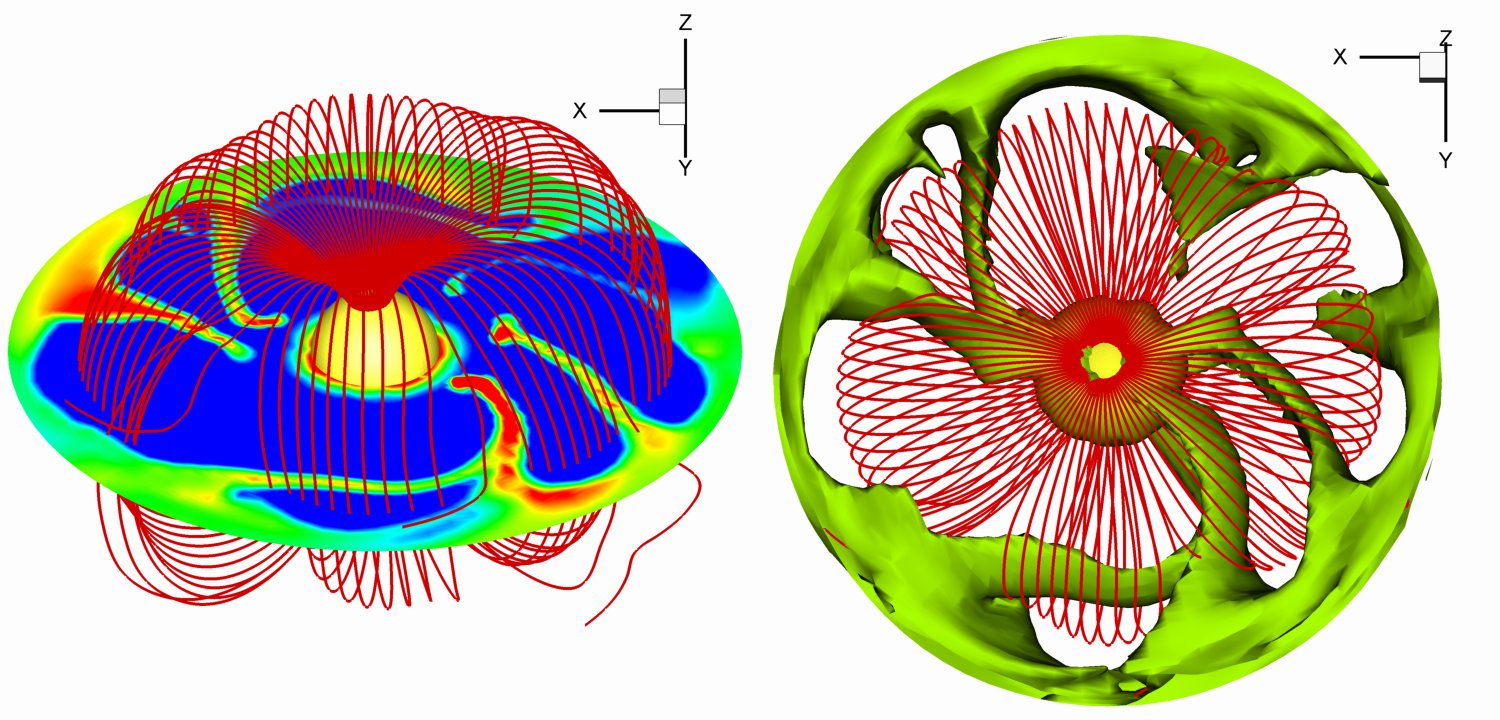}{push-aside}{{\it Left panel:} A tongue of gas, shown by density contours in the equatorial plane, pushing aside magnetic field lines on its way to the star. Note that the ``hole'' in the magnetosphere is not artificially depicted -- the field lines start out uniformly spaced on the star's surface, and twist aside around the tongue. {\it Right panel:} A constant density surface at the same instant of time as in the left panel. The grid resolution is $144\times61^2$.}

\fig{3d-main} shows two views of the accretion flow around the star through instabilities (top row) and two views of a magnetospheric (funnel) flow (bottom row). The growth of unstable perturbations at the disk-magnetosphere boundary results in penetration of the magnetosphere by the disk matter, in the form of tongues of gas travelling through the equatorial plane. Matter energy density dominates inside the tongues. When they come closer to the star, they encounter a stronger magnetic field, which stops their equatorial motion. At this point the tongues turn into miniature funnel-like flows following the field lines, which gives them a characteristic wishbone shape. They deposit matter much closer to the star's equator than true funnel flows do.

The tongues are tall and thin, as opposed to the funnels which are flat and wide. This is because the tongues penetrate the magnetosphere by prying the field lines aside (\fig{push-aside}), since this is energetically more favourable than bending the field lines inward. This is a standard feature of the interchange instability --- if a heavy fluid is supported against gravity by a light fluid, then at the boundary between the fluids, fluid elements on either side displace those on the other side. Since the magnetic field is frozen into the matter, the field lines are also pushed aside in the process. This interchange process continues, producing fingers of each fluid penetrating into the other (see, e.g., Arons \& Lea 1976a, Wang \& Robertson 1985). The tension of the magnetic field lines suppresses the interchange process in the direction parallel to the field. This increases the characteristic wavelength of deformation of the boundary in that direction (Chandrasekhar 1961). Hence, the tongues are narrow in the direction perpendicular to the field, i.e., the azimuthal direction, and broader parallel to the field, i.e., the vertical direction (Stone \& Gardiner 2007a,b).

\fig{push-aside} shows that the magnetosphere of the star is strongly disturbed by the penetrating tongues which push the field lines aside. To conserve angular momentum, the angular velocity of the gas in the tongues increases as it moves inwards, making it rotate faster than the inner disk matter, causing the tongues to curve to their right. This tongue shape allows tongues that move faster to merge with more slowly moving tongues. As a result, the total number of tongues at any given time is of the order of a few. Such merging of Rayleigh-Taylor fingers has been observed in earlier numerical simulations (Wang \& Robertson 1985). The merging and subsequent growth of the tongues occurs on the inner-disk dynamical timescale. The number of tongues we see is of the order of a few.

\fignar{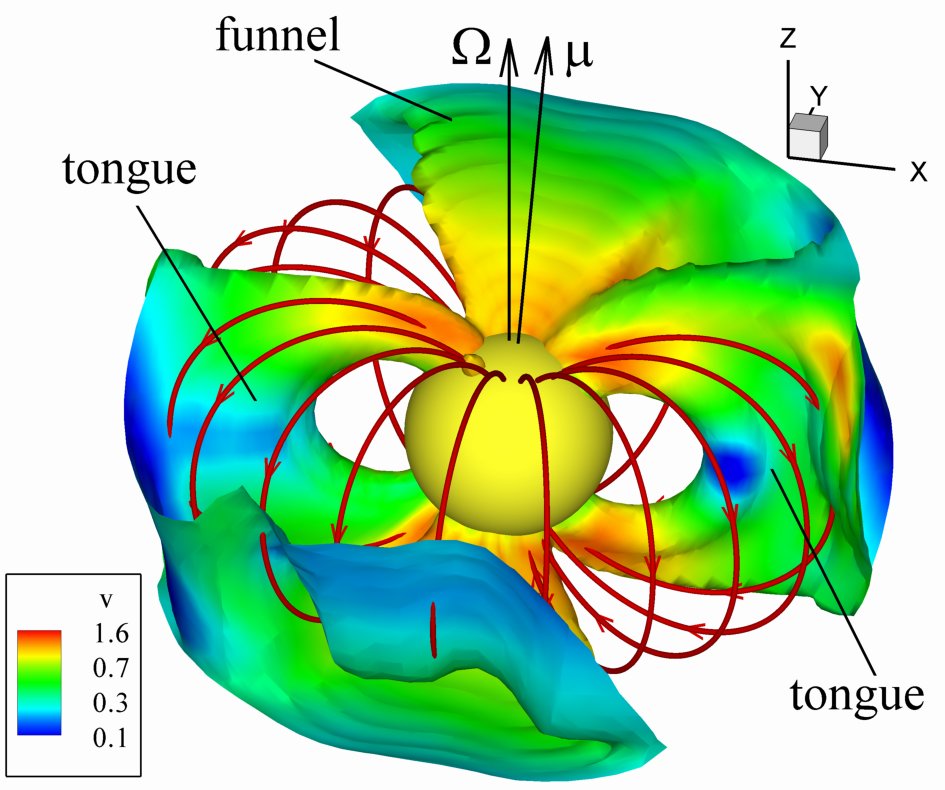}{ro-v}{Cutaway view of the region around the star showing the matter velocity profile in the funnels and tongues, in a reference frame rotating with the star. A constant density surface is shown, overlaid with velocity contours. $\Omega$ and $\mu$ are the star's rotation and magnetic axes.}

\figwide{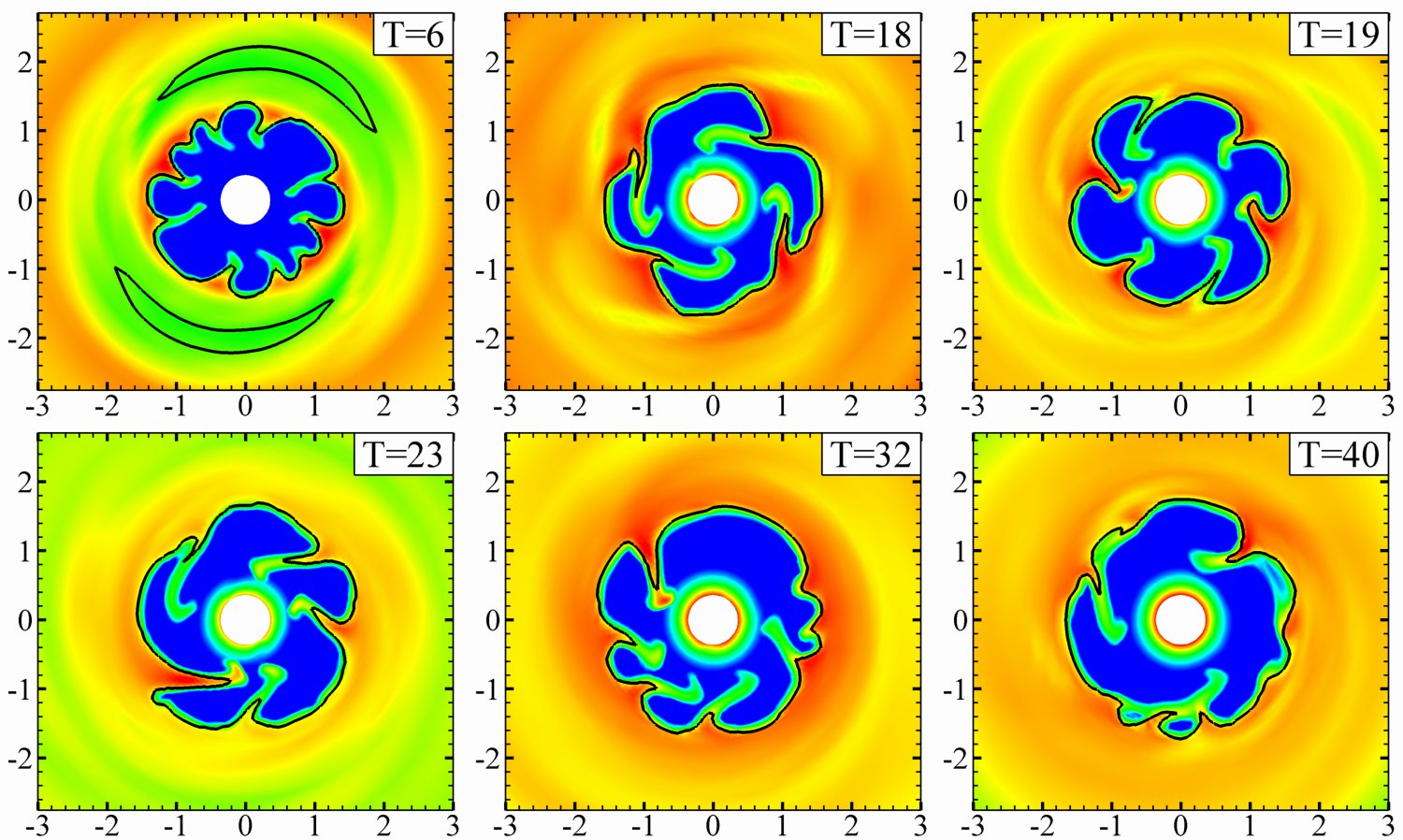}{xy-main}{Equatorial slices of the circumstellar region for our main case. The colors represent plasma density contours, ranging from red (highest) to blue (lowest). The black line is the $\beta=1$ line.}

The matter in the tongues is accelerated by the gravitational field of the star, so that its velocity is higher near the star. The density and velocity distribution in the tongues is similar to that in the funnel streams, as \fig{ro-v} shows. It is also seen that the funnels and tongues are not mutually exclusive. We discuss this last point in more detail in \S\ref{sec:alpha-dep} and \S\ref{sec:instab-regime}.

\fig{xy-main} shows equatorial slices of the circumstellar region at various times. The density enhancements which result in the formation of the tongues can be seen at the bases of the tongues. The number of tongues changes with time of its own accord, without any artifically introduced perturbation.

To estimate the number and positions of the tongues at different times, we plot the plasma density in the equatorial plane along circles of radius $R$ centered at the star, as a function of time (\fig{tongues-main}). The density enhancements show the positions of the tongues. The dashed vertical lines show how many tongues cross a certain radius at a specific time. The top panel shows the evolution of the tongues at $R=1.4$. This is close to the inner radius of the disk just before the start of the instability. This is the place where the number of tongues is the largest (usually 4-7). The tongues are dense at this radius. The middle panel shows that closer to the star, at $R=1$ there are usually 3-4 tongues, and their density is smaller. Even closer to the star, at $R=0.6$ (lower panel), the tongues become weaker and their number decreases to 1-3. This is because close to the star, many of the tongues leave the equatorial plane and travel along the magnetic field lines to the star.

This plot also shows us the azimuthal positions $\phi$ of the tongues as a function of time, giving a rough idea of their angular velocity. The azimuthal position $\phi$ is measured in the coordinate system rotating with the star. We can see that the tongues move and change shape on the dynamical timescale in the inner disk region. The number of tongues varies between about 2 and 7.

\figwide{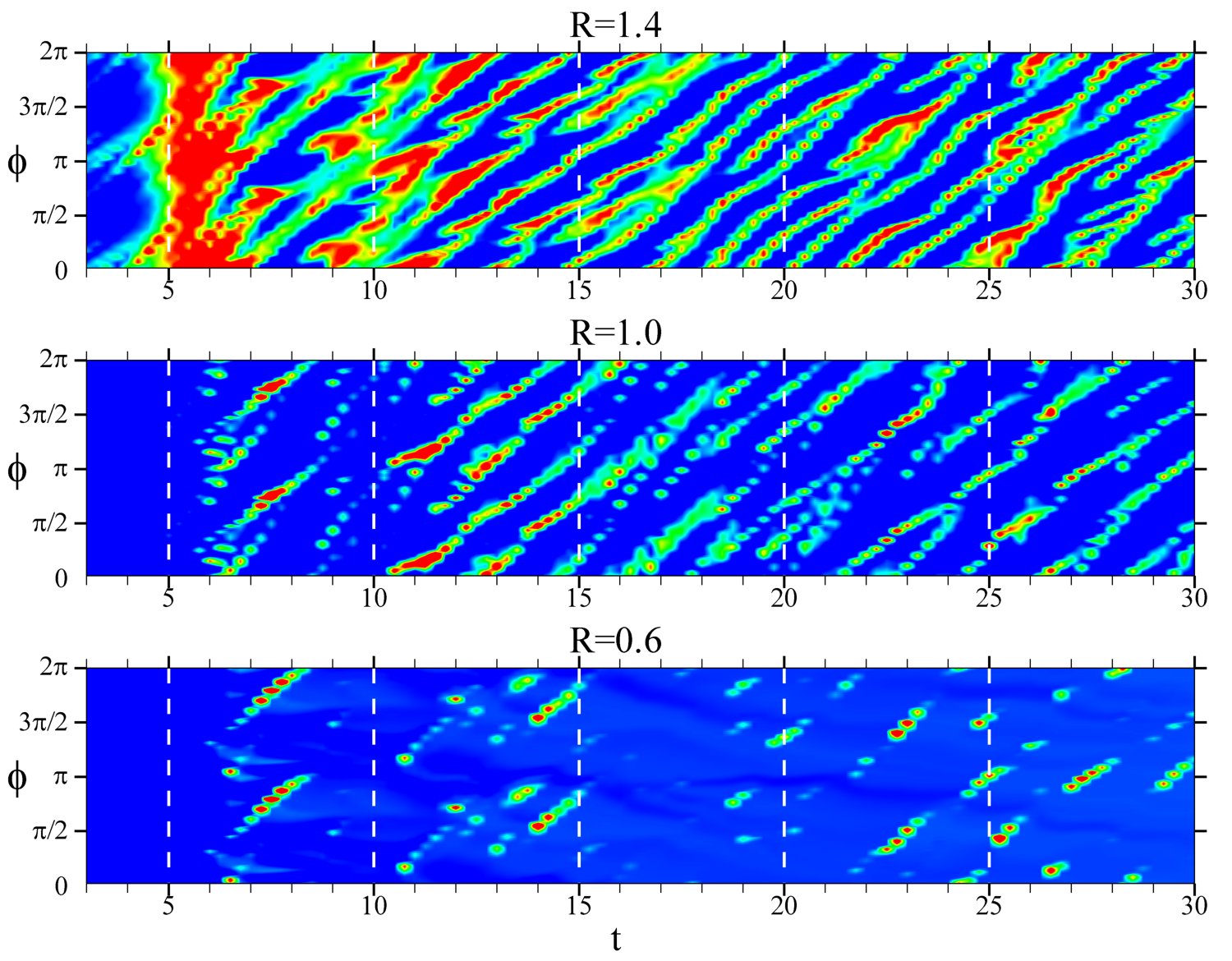}{tongues-main}{Azimuthal plasma density distribution in the equatorial plane as a function of time at various radii. The colors represent plasma density levels, ranging from 0.01 (deep blue) to 1 (red).}

\fig{spots-pictures} shows the hot spots on the star's surface for our main case at different times. We see that the spots are different from pure funnel-flow hot spots (Romanova et al. 2004, Kulkarni \& Romanova 2005), and are significantly different from the simple polar-cap shape that is frequently assumed. Each tongue creates its own hot spots when it reaches the star's surface. Therefore, the shape, intensity, number and position of the spots change on the inner-disk dynamical timescale. As noted earlier in this section, since the density and velocity of the matter in the tongues is comparable to that in the funnel streams (\fig{ro-v}), the energy flux from the hot spots created by the tongues is also comparable to that from funnel-stream hotspots in other, stable cases. \fig{lcurve} compares the typical lightcurves during accretion through funnels and through instability. The lightcurve is usually very chaotic during accretion through instability, and shows no definite periodicity. Sometimes, however, a certain number of tongues dominates, and the lightcurve may show quasi-periodicity. Also, as mentioned above, funnels and tongues coexist for certain ranges of parameter values. In this intermediate regime, a mixed lightcurve with both periodic and chaotic components is expected. We plan to discuss this in more detail in a future paper.

\figwide{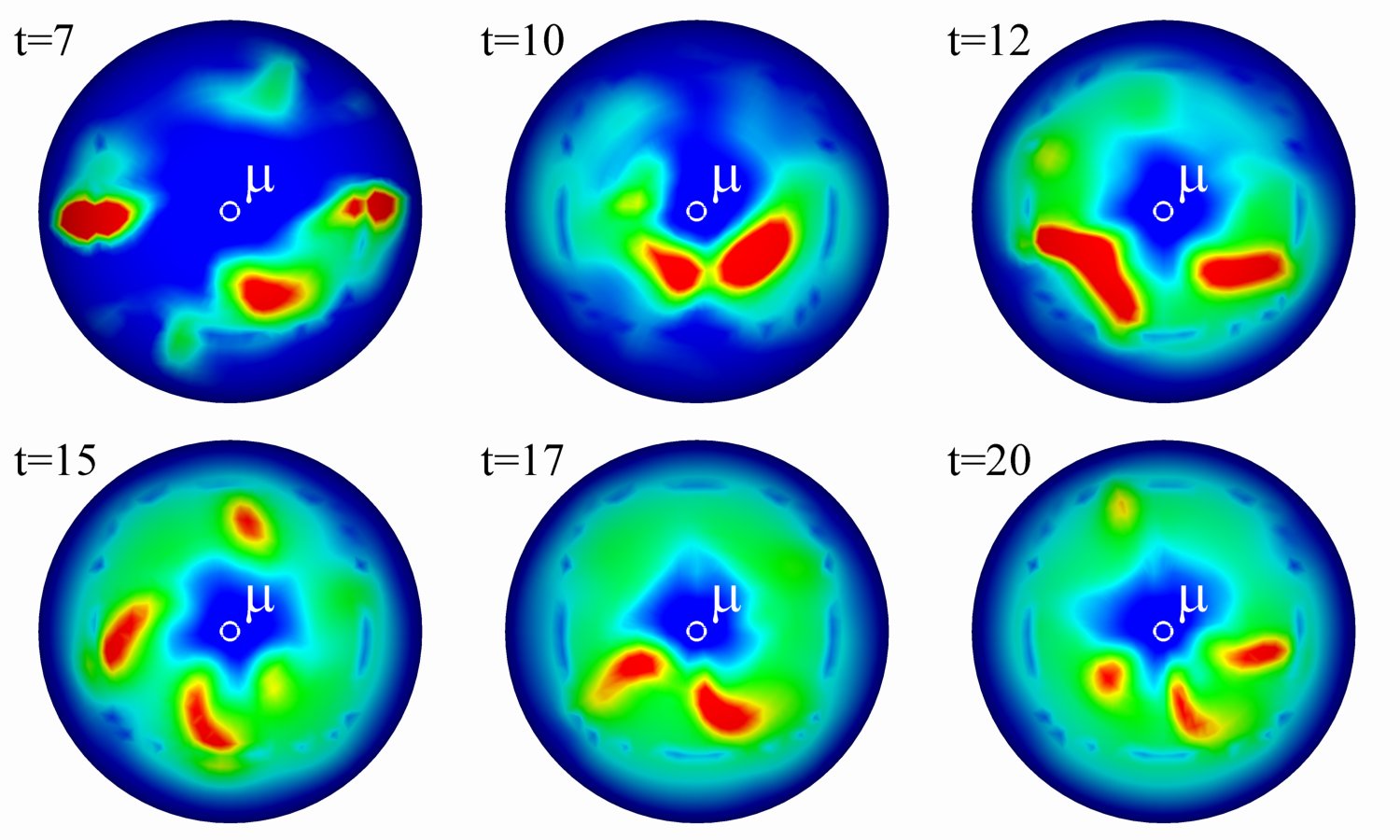}{spots-pictures}{Hot spots on the star's surface at various times for our main case. The colors represent contours of the matter flux onto the star's surface, ranging from 0.01 (deep blue) to 3 (red).}

\figwide{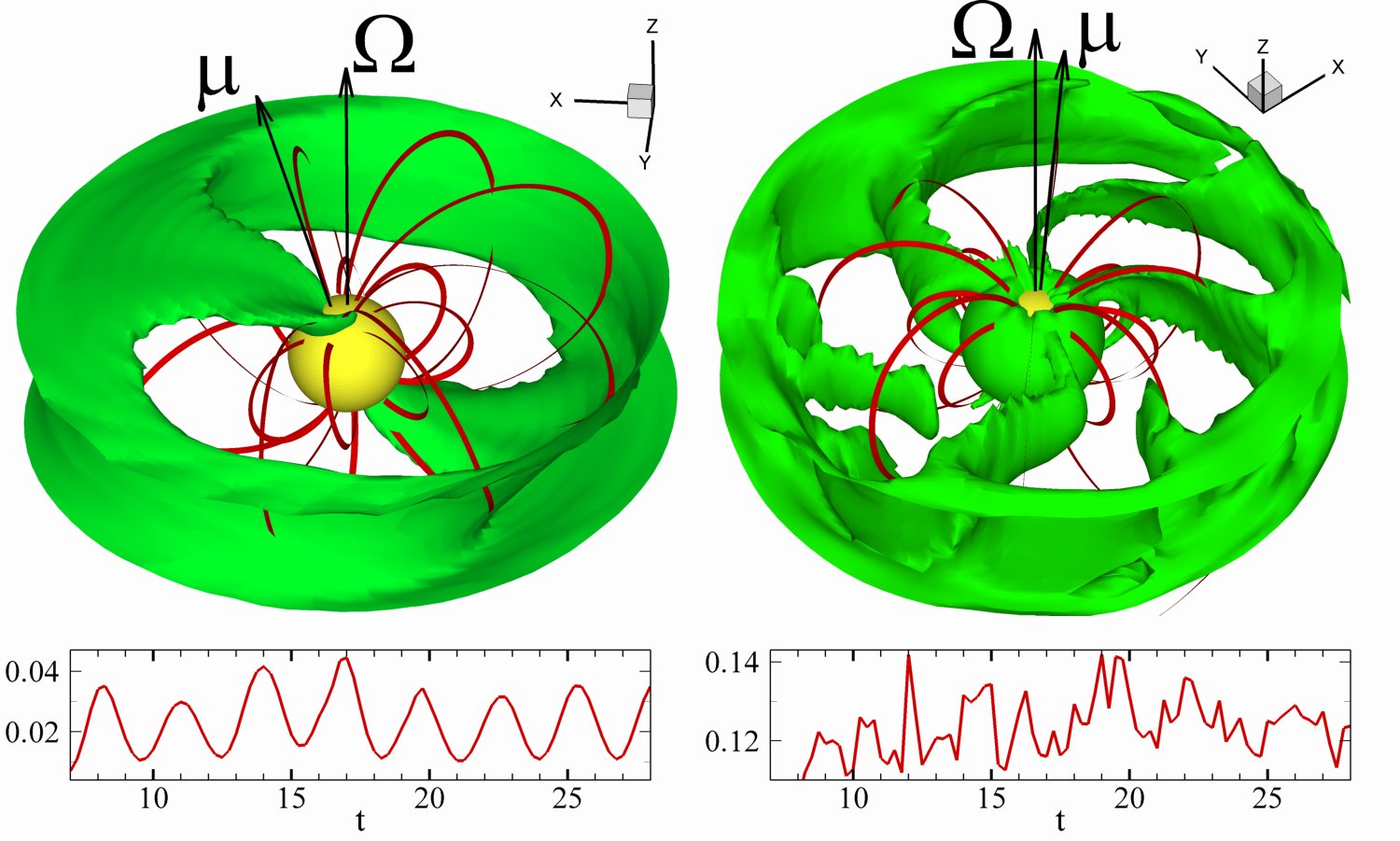}{lcurve}{Accretion flow and corresponding lightcurves during funnel accretion (left panel) and during accretion through instabilities (right panel). The grid resolution is $144\times61^2$.}

In the following subsections, we investigate the dependence of the instability on different parameters, varying one parameter at a time and choosing the same values for the other parameters as in the main case.

\subsection{Dependence on the Accretion Rate}
\label{sec:alpha-dep}

The accretion rate $\dot M$ in our code is regulated by the viscosity coefficient $\nu$ which is proportional to the $\alpha$-parameter (Shakura \& Sunyaev 1973). The radial velocity of inward flow in the disk at a distance $r$ from the star is $v_r \sim \nu/r \sim \alpha c_s h/r$, where $h$ is the thickness of the disk and $c_s$ is the sound speed. Thus the accretion rate through the disk is approximately proportional to $\alpha$: $\dot M\approx 4\pi r h \rho v_r \sim \alpha$. The accretion rate is also proportional to the fiducial density $\tilde{\rho}$ in the disk, but we keep $\tilde{\rho}$ fixed in all our runs (except in test runs that we describe at the end of this subsection), giving us the same initial density distribution in all runs.

We performed simulations for a wide range of the parameter $\alpha$, from very small to relatively large, $\alpha= 0.02, 0.03, 0.04, 0.06, 0.08, 0.1, 0.2, 0.3$. We find that at very small $\alpha$ $(\leq 0.03)$, the instability does not appear. At larger $\alpha$'s, the instability appears, and when $\alpha$ is increased, the instability starts earlier and more matter accretes through it. \fig{xy-alpha} shows equatorial slices of the plasma density distribution at different $\alpha$. One can see that there are no tongues at $\alpha=0.02$. The tongues are quite weak at $\alpha=0.04$, but much stronger at larger $\alpha$, when more matter comes to the inner region of the disk, and the plasma density in the inner region of the disk is higher than in the low-$\alpha$ cases. This shows that increased accumulation of mass at the inner edge of the disk leads to enhancement of the instability, producing tongues that propagate deeper into the magnetosphere of the star. We should note that in spite of different conditions at the inner region of the disk (much higher density at larger $\alpha$), the number and behaviour of the tongues is approximately the same in all cases.

\figwide{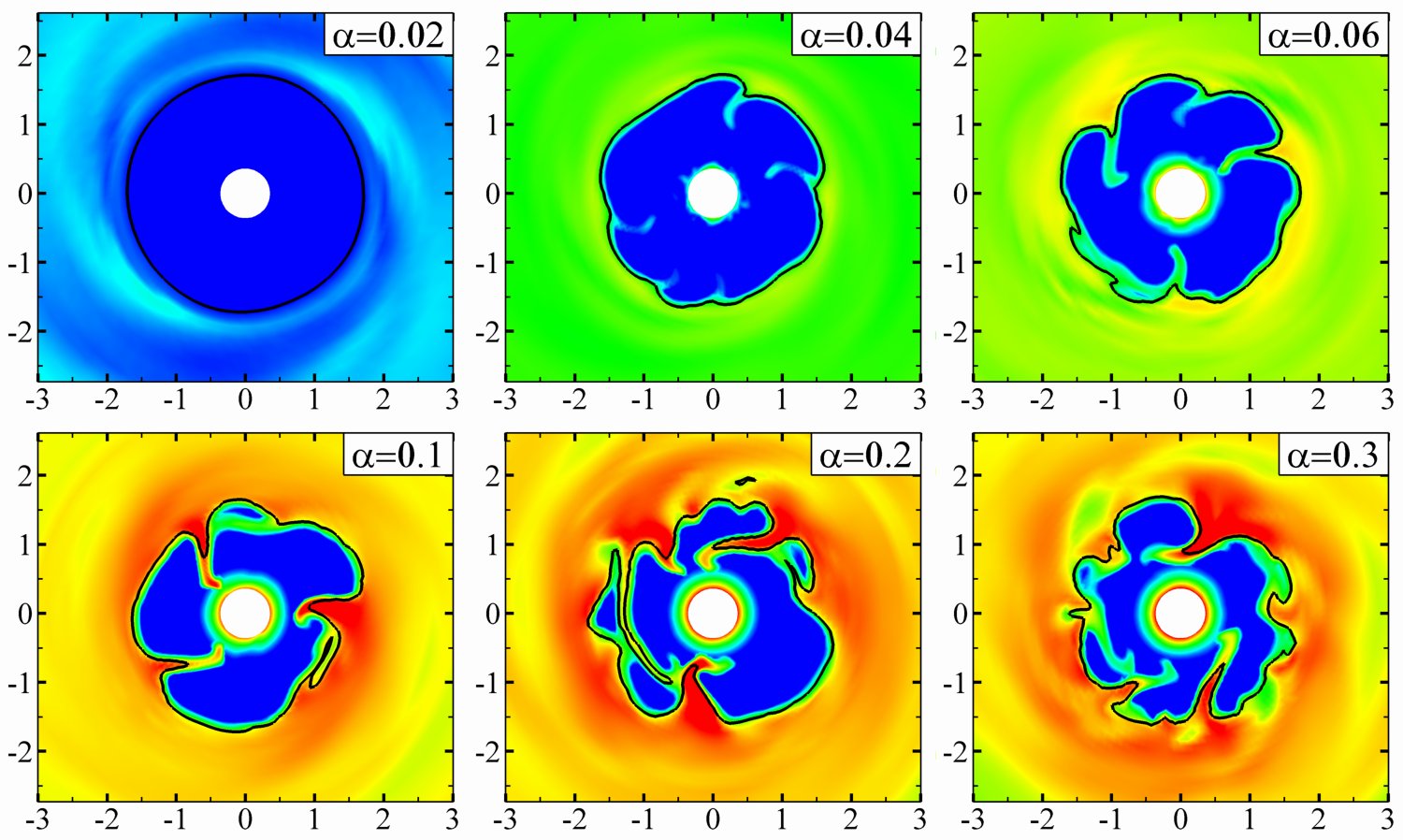}{xy-alpha}{Plasma density distribution in the equatorial plane for different $\alpha$ values. The colors represent plasma density contours, ranging from red (highest) to blue (lowest). The black line is the $\beta=1$ line.}

\fig{xz-alpha} shows the density distribution in the $\mu-\Omega$ plane. One can see that at $\alpha=0.02$, the accretion is entirely through magnetospheric funnel streams. At $\alpha=0.04$ and $\alpha=0.06$, a significant amount of matter accretes through the funnel streams, though some accretes through instabilities. We call this the intermediate regime of accretion. The bottom row shows the most unstable cases, where most of the matter flows through the tongues. For the unstable cases, what we see inside the magnetosphere in this figure is cross sections of the tongues.

\figwide{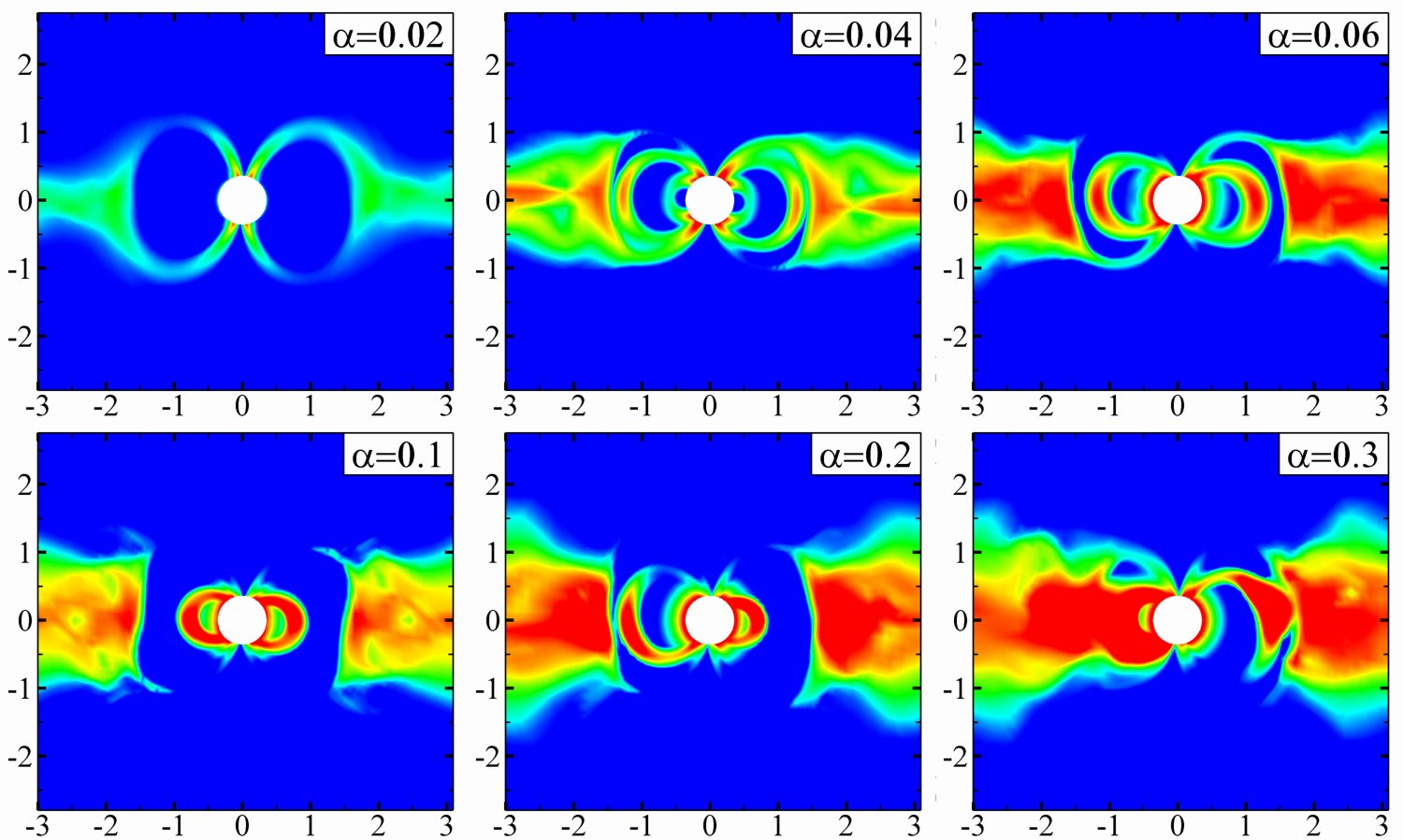}{xz-alpha}{Density distribution in the $\mu-\Omega$ plane for different $\alpha$ values. The colors represent plasma density contours, ranging from red (highest) to blue (lowest).}

The accretion rate onto the star's surface is higher during accretion through instability, as \fig{pm-alpha} shows. We see that the accretion rate increases with increasing $\alpha$. This is mainly due to increase in the amount of matter transported inwards by the accretion disk.
The higher accretion rate is accompanied by a higher angular momentum flux (\fig{puf-alpha}).

\fignar{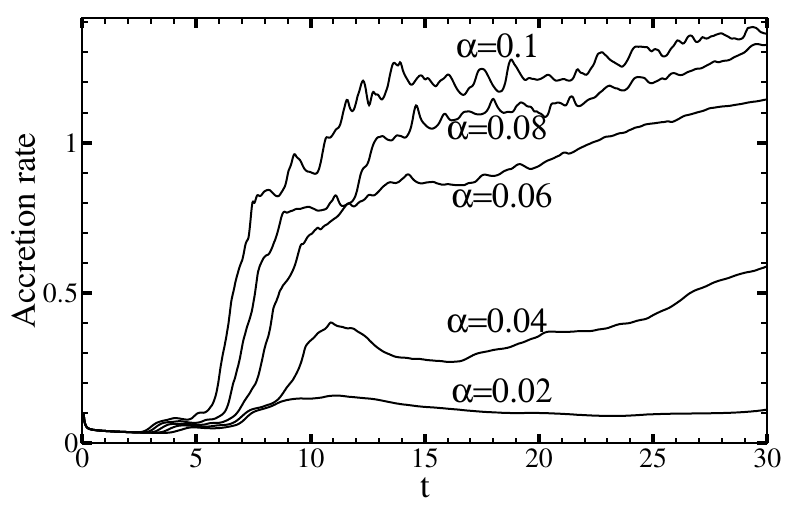}{pm-alpha}{Variation of the accretion rate onto the star's surface with time, for different values of the $\alpha$-viscosity.}

\fignar{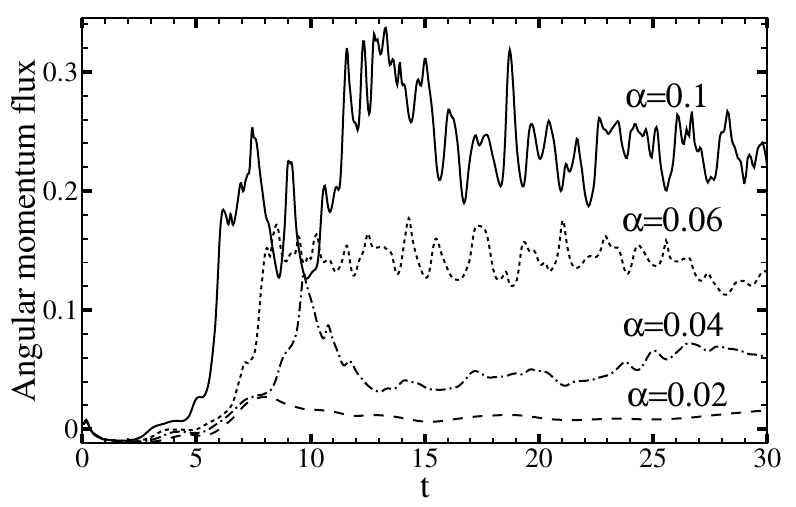}{puf-alpha}{Variation with time of the angular momentum flux onto the star about the star's rotational axis, for different values of the $\alpha$-viscosity.}

The position and number of the hot spots on the star's surface varies with time. To find the average location of the spots, we calculated the matter flux integrated over magnetic longitude, $\dot M(\theta_m)$, as a function of magnetic latitude $\theta_m$. $\dot M(\theta_m)$ is defined such that the total accretion rate onto the star, $\dot M = \int \dot M(\theta_m) d\theta_m$. \fig{pm-lat-time} shows the evolution of $\dot M(\theta_m)$ with time. The top panel shows that at large $\alpha$, when the instability is strong, the hot spots are located at the mid-latitudes, being brightest at $35^\circ < \theta_m < 65^\circ$. The bottom panel shows that at very small $\alpha$, when matter accretes through magnetospheric funnels, the spots are located at much higher latitudes, $60^\circ < \theta_m < 75^\circ$. In the intermediate case, $\alpha=0.04$ (middle panel) both types of accretion are present and the plots reflect hot spots produced both by magnetospheric streams and by instabilities.

\fignar{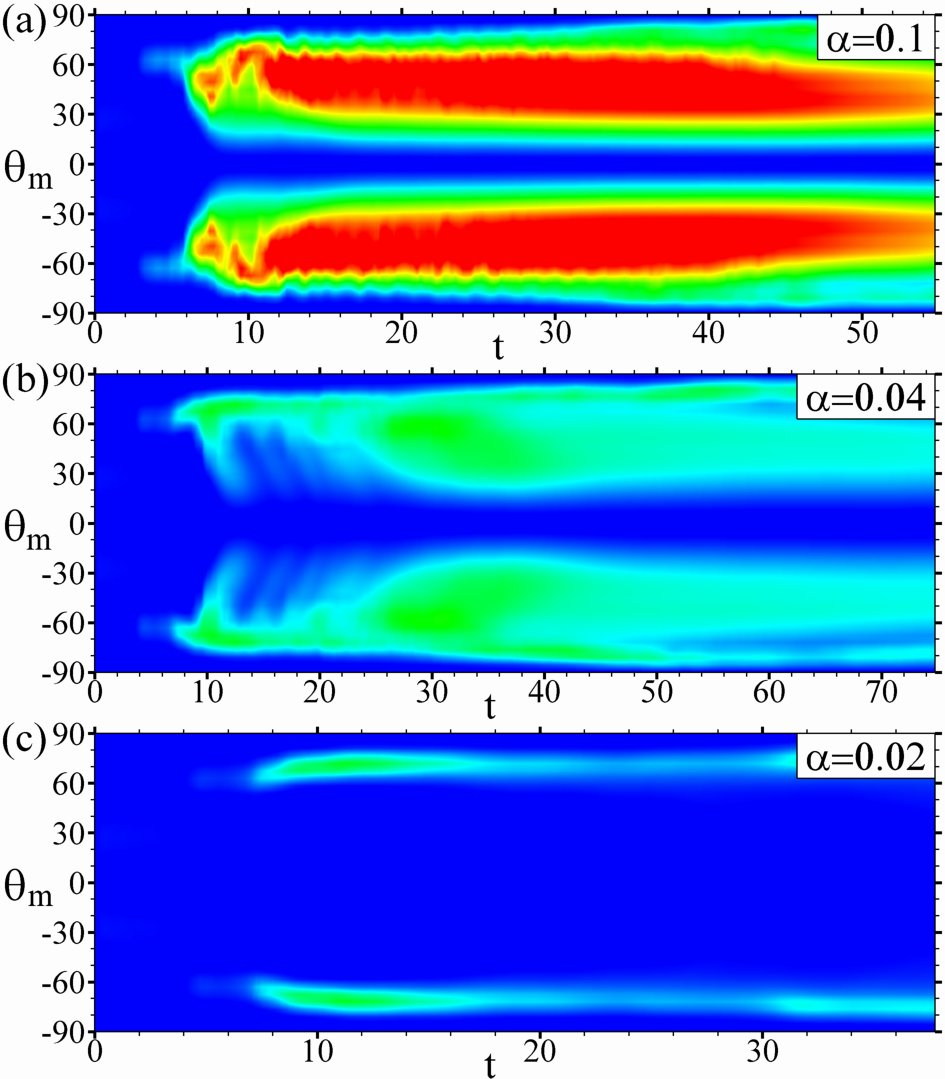}{pm-lat-time}{Distribution of matter flux $\dot M(\theta_m)$ onto the star's surface integrated over magnetic longitude, as a function of magnetic latitude $\theta_m$ and time. The colors represent contours of the integrated matter flux, ranging from 1 (deep blue) to 30 (red).}

\fig{pm-lat-alpha} shows the $\theta_m$-dependence of $\dot M(\theta_m)$ at $t=25$ for different $\alpha$. The plot shows that at $\alpha=0.02$ and $\alpha=0.03$ the maximum of matter flux is located at $\theta_m\approx 70^\circ$ with the half-width of the peak $\approx 75^\circ - 65^\circ = 10^\circ$. For $\alpha=0.06-0.1$, the maximum is at much lower latitudes, $\theta_m\approx 50^\circ$ with half-width $\approx 70^\circ - 25^\circ = 45^\circ$. It is surprising to see that at the largest viscosity coefficients, $\alpha=0.2$ and $0.3$, the hot spots do not move closer to equator, but have a maximum at $\theta_m\approx 50^\circ$, like for $\alpha=0.1$.

\fignar{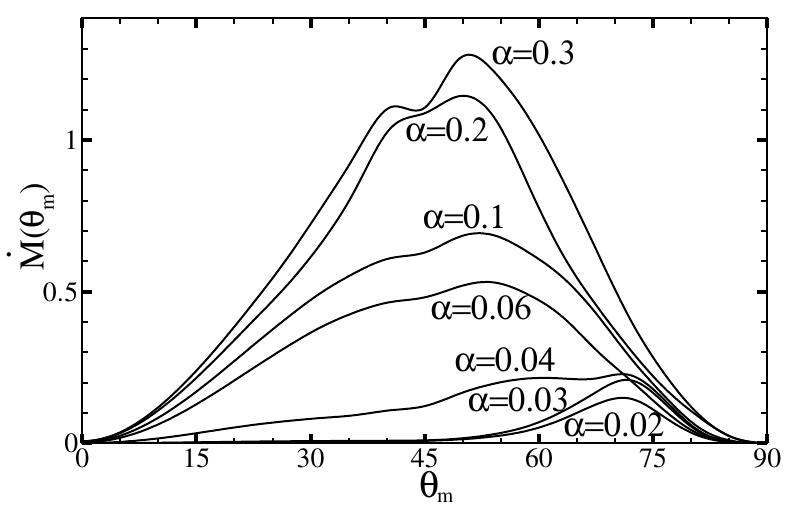}{pm-lat-alpha}{Dependence of the longitude-integrated matter flux on the magnetic latitude at $t=25$ for different $\alpha$.}

To eliminate the possibility that $\alpha$-viscosity is directly responsible for the instability, we performed test simulation runs at $\tilde{\rho}=2$ and $\alpha=0.02$, and found that the instability develops in this case, although, as noted earlier, the run at $\tilde{\rho}=1$ and $\alpha=0.02$ is stable. Similarly, in a run with $\tilde{\rho}=0.5$ and $\alpha=0.04$, no instability is observed, although for $\tilde{\rho}=1$ and $\alpha=0.04$, the instability exists. This and a few other test runs have shown that it is really the matter flux, and not $\alpha$ itself, that is responsible for the onset of the instability.

This subsection shows that the accretion rate (controlled by $\alpha$) is one of the important factors determining whether the matter flow is stable (with magnetospheric funnel flows) or unstable, where most of the matter flows through equatorial instabilities.

\subsection{Dependence on the magnetic field strength and the star's rotation rate}
\label{sec:dip-rc-dep}
In qualitative terms, the instability depends on the effective gravitational force (i.e., the difference between the gravitational and centrifugal forces) acting on the inner disk matter. The star's magnetosphere forces the matter at the inner disk boundary to approximately corotate with the star. So if the magnetospheric radius $r_m$ is significantly smaller than the corotation radius $r_{cor}$, then the inner disk matter is appreciably slowed down by the magnetosphere, and the effective gravitational force is large. The effect of this most clearly seen when comparing runs with different magnetic fields or with different stellar rotation rates.

First, starting from our main case, with parameters $\alpha=0.1$, $\tilde{\rho}=1$, $P=3$, $r_{cor}=2$ and $\Theta=5^\circ$, we varied the magnetic moment $\mu$ of the star. We observed that the instability appears at a wide range of $\mu$, from $\mu=0.2$ to $\mu=8$. \fig{xy-dip} shows that the small $\mu$, $r_m$ is very small, and also the typical mode number $m=2$. At very large $\mu$ ($\mu=8$), $r_m \approx 1.8-2.2$ becomes comparable to $r_{cor}=2$, and the star is in the propeller regime, so that the effective gravity is very small or even radially outwards, and is unable to drive the instability. Thus, increasing the magnetic field leads to suppression of the instability by decreasing the effective gravity.

In another set of runs, we started from our main case and fixed $\mu=2$ (which fixes $r_m$), but varied the rotation period $P$ of the star. We observed that at smaller $P$, the instability becomes weaker and is finally suppressed, again because decreasing the period decreases $r_{cor}$, bringing it closer to $r_m$, taking the star into the propeller regime and reducing the effective gravity. We discuss the dependence of the instability on the star's magnetic field and rotation rate in more detail in \S\ref{sec:instab-regime}.

\figwide{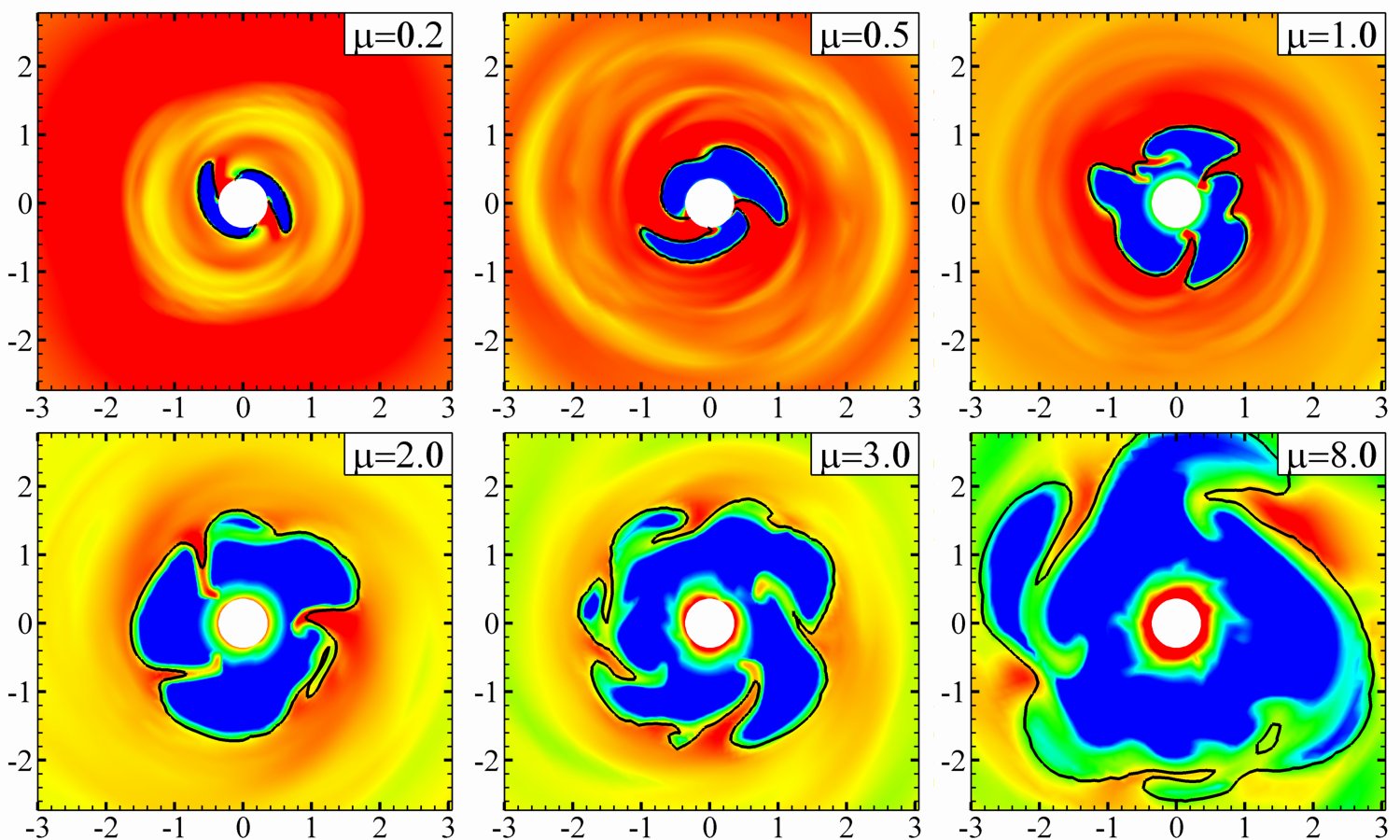}{xy-dip}{Density distribution in the equatorial plane for different magnetic moments $\mu$. The colors represent plasma density contours, ranging from red (highest) to blue (lowest). The black line is the $\beta=1$ line.}

\subsection{Dependence on the misalignment angle}
\fig{xy-theta} shows the equatorial matter flow at various misalignment angles $\Theta$, all other parameters being the same as in our main case. We find that the instability shuts off for $\Theta \gtrsim 30^\circ$. The reason for this is that for large misalignment angles ($\Theta \gtrsim 30^\circ$), the magnetic poles are closer to the disk plane. Therefore, the gravitational energy barrier that the gas in the inner disk region has to overcome in order to form funnel flows is reduced, making funnel flows energetically more favourable. The matter seen inside the magnetosphere for $\Theta=60^\circ$ is part of the warped funnel stream that crosses the disk plane. We also find that for $\Theta=30^\circ$, the $m=2$ mode (two tongues) usually dominates.

\fig{pm-lat-theta-rot} shows the accretion rate $\dot M(\theta)$ onto the star's surface as a function of rotational latitude $\theta$. We see that when the accretion is through instability ($\Theta \leq 30^\circ$), most of the matter accretes onto the mid-latitude ($\theta \sim 50^\circ$) region of the star, {\it independent of the star's misalignment angle.}

\subsection{Dependence on the grid resolution}
We found that the azimuthal extent of each tongue is much larger than the size of our grid cells. This indicates that the instability is not an artefact of the coarseness of the grid. Nevertheless, to eliminate that possibility, we performed test simulations at higher grid resolutions.
\fig{xy-grid} shows equatorial slices of the region near the star at various grid resolutions. The instability exists at all the resolutions we tested, including a test run at resolution $216\times91^2$ (not shown here). The number and behaviour of the tongues is similar in all these cases. However, with the finer grid, the tongues are thinner on an average. The accretion rate onto the star at the finest grid is about 20\% larger than that with the coarsest grid, which may be related to the smaller numerical diffusivity at finer grids, and the resultant better coupling between the outer magnetosphere and the disk, which leads to a higher accretion rate.

\figwide{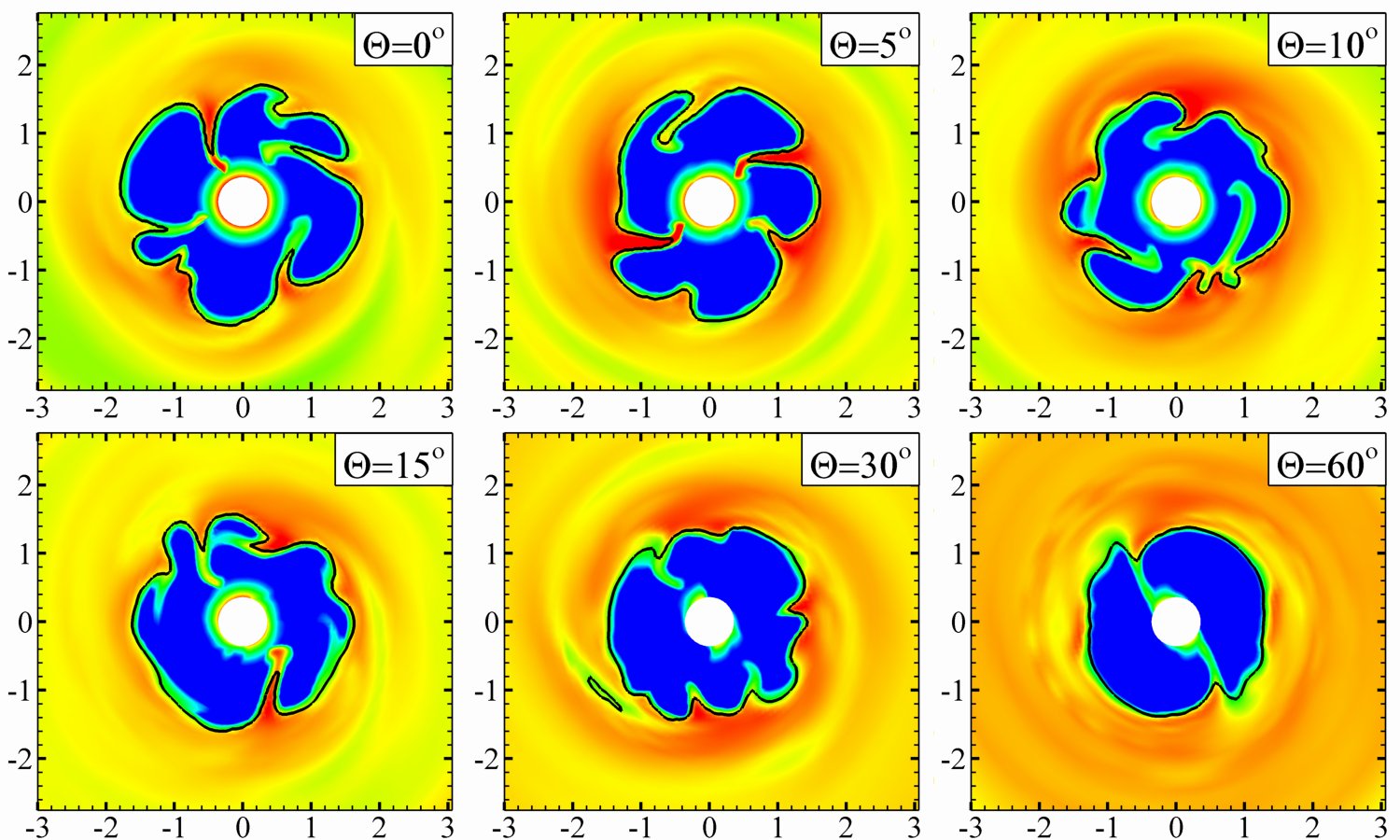}{xy-theta}{Equatorial slices of the circumstellar region at various misalignment angles. The colors represent plasma density contours, ranging from red (highest) to blue (lowest). The black line is the $\beta=1$ line.}

\fignar{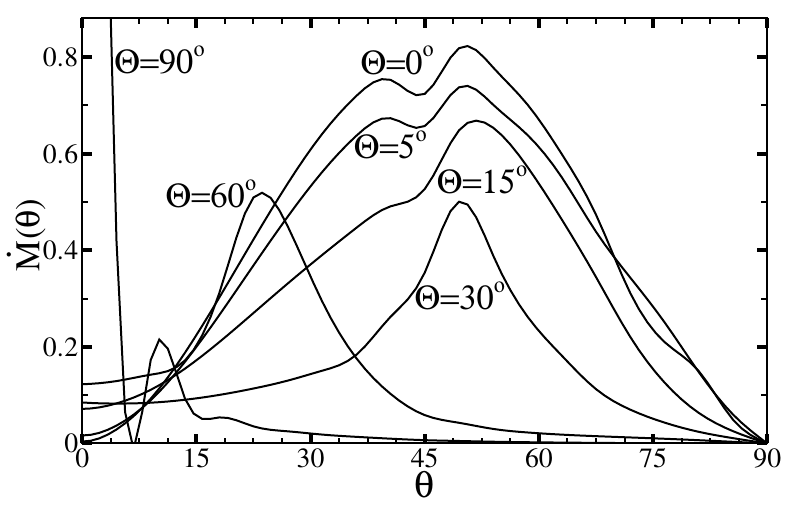}{pm-lat-theta-rot}{Dependence of the longitude-integrated matter flux $\dot M(\theta)$ on the rotational latitude $\theta$ at t=32 for different $\Theta$.}

\figwide{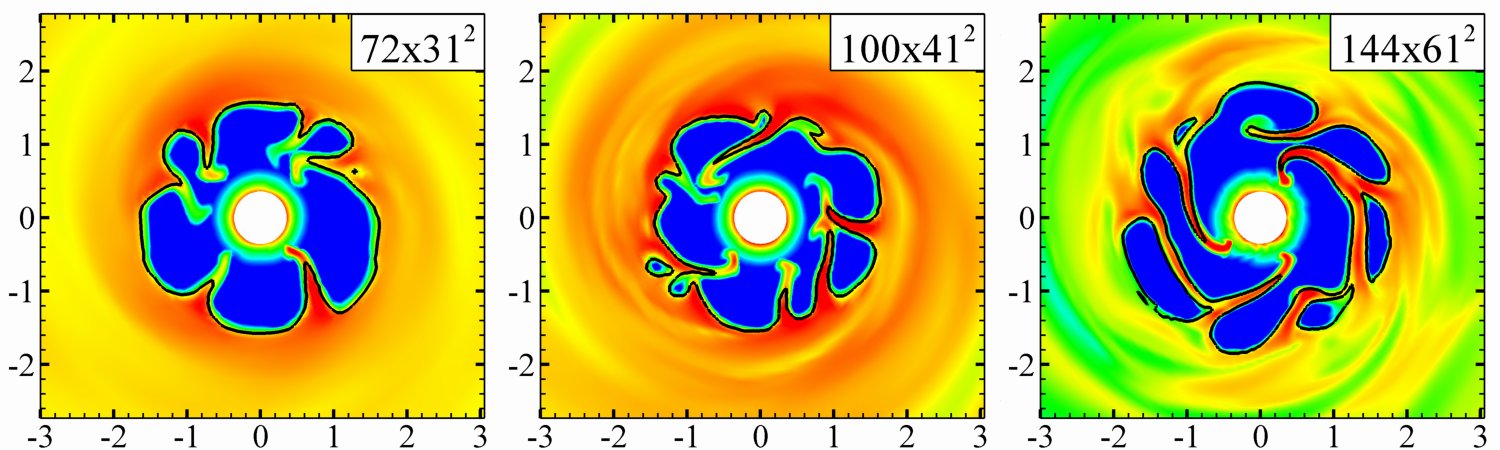}{xy-grid}{Equatorial slices of the circumstellar region at various grid resolutions and times. The colors represent plasma density contours, ranging from red (highest) to blue (lowest). The black line is the $\beta=1$ line. The grid resolutions are specified as $N_r\times N^2$, where $N_r$ and $N$ are the numbers of grid cells in the radial and angular directions respectively in each of the six zones of the cubed-sphere grid.}

\subsection{Possible perturbation mechanisms}
The Rayleigh-Taylor or Kelvin-Helmholtz instabilities will only manifest themselves in a potentially unstable layer between two liquids if some perturbation of density, pressure or velocity occurs in the layer. There are different mechanisms of perturbation which are expected in real accretion disks. First of all, the matter in the disk is never perfectly homogeneous, and natural density and pressure inhomogeneities may act as perturbations. Also, if the magnetic and rotational axes are not aligned, there will always be some density enhancement near the disk foot-points of the funnel streams (Romanova et al. 2003, 2004). This would be a constant source of inhomogeneity in the disk. Our simulations have shown that even at small misalignment angles, $\Theta\sim 5^\circ$, the misalignment leads to an inhomogeneous density distribution in the disk, with two oppositely oriented density enhancements or spiral waves. The effect is even stronger at larger misalignment angles, $\Theta\sim 10-30^\circ$. Another source of perturbation is associated with the magnetic field lines which are trapped inside the inner regions of the disk and are azimuthally wrapped by the disk matter. This leads to increase of magnetic energy in some parts of the disk and to partial expulsion of matter from these regions, and thus to inhomogeneous distribution of matter. This mechanism is expected to operate in real astronomical objects as well.

Concerning the role of the grid, it is unlikely, as noted above, that the discrete nature of the grid by itself leads to perturbations. But another perturbing element is the {\it boundary} between the sectors of the cubed sphere grid. Four of these boundaries cross the disk. They produce initial density and pressure perturbations at the 5\% level near the disk-magnetosphere boundary, and at even larger levels at larger distances from the star, where the grid is coarser. At later times these perturbations become less important. So at early times in the simulations, this boundary effect is the most important contributor to the perturbations. That is why we often see four tongues initially. However, at later times, we often observe anywhere between 2 and 7 tongues, which shows that there is no direct influence of these boundaries on the perturbations at later times.

\figwide{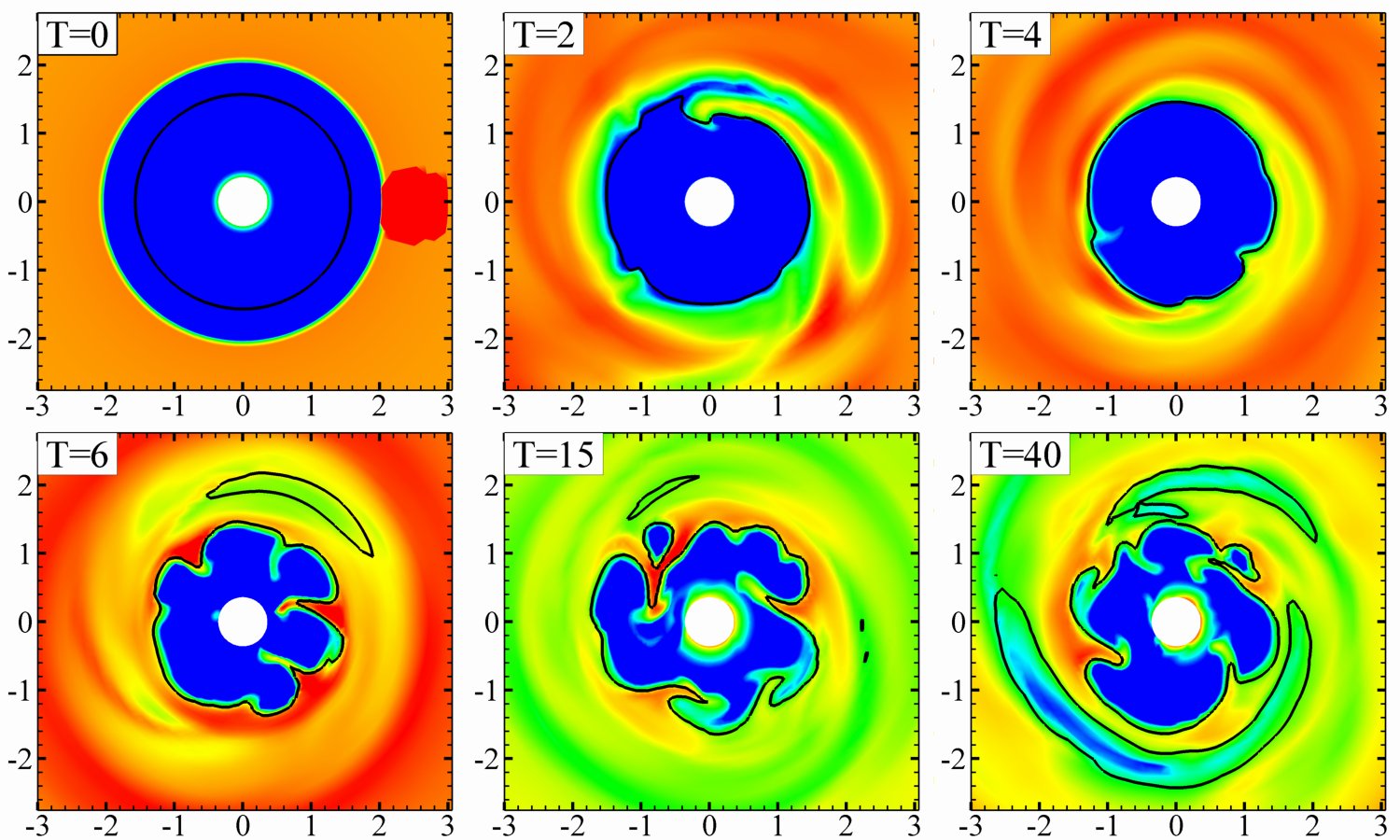}{xy-blob}{Equatorial slices of the circumstellar region at various times for our main case, with an artificially introduced density enhancement at the inner edge of the disk at t=0. The colors represent plasma density contours, ranging from red (highest) to blue (lowest). The black line is the $\beta=1$ line.}

To check the importance of the boundaries between the grid sectors in the generation of perturbations, we introduced a perturbation in the form of a density enhancement at the inner edge of the disk at the beginning of the simulation. At a certain location in the inner disk region, we chose a sphere of diameter equal to the initial thickness of the disk, and increased the plasma density in the sphere by a factor of 10. \fig{xy-blob} shows the temporal evolution of the instability in this case. We see that the perturbation generated by the blob leads to the formation of more than four tongues right from the beginning of the simulation. This shows that the initial perturbations by the boundaries of the grid and by the blob lead to very similar subsequent evolution with similar numbers of unstable tongues.

Another issue is whether the growth of the tongues is the effect of the numerical diffusivity of the code. We do not think this is the case, since the tongue growth occurs on the inner-disk dynamical timescale, which is much shorter than the diffusive timescale at this distance. Also, the numerical diffusivity decreases when the grid resolution is increased, but the instability still exists at higher resolutions.

\section{Empirical conditions for the existence of the instability}
\label{sec:instab-regime}
To investigate in more detail the parameter ranges over which the instability appears, we performed multiple simulation runs for a variety of values of $\alpha$ and $P$, for two values of the magnetic dipole moment, $\mu=2$ and $\mu=0.5$, at misalignment angle $\Theta=5^\circ$. \fig{stab-nonstab} shows the resulting regimes of stable and unstable accretion. The top panel shows that, as noted before, a high accretion rate through the disk and slow rotation of the star favour the instability. The bottom panel shows the stable and unstable regimes in the $\dot{M_*}-P$ plane, where $\dot{M_*}$ is the accretion rate onto the surface of the star. Here, the boundary between the regimes has a much weaker dependence on the rotation period. This is probably because, as mentioned in \S\ref{sec:dip-rc-dep}, increasing the star's rotation rate takes the star closer to the propeller regime, lowering the disk density. The combination of the reduced disk density and higher $\alpha$ possibly produces an almost constant accretion rate onto the star's surface.

It is to be noted that, as described in \S\ref{sec:alpha-dep}, the transition between the stable and unstable regimes is not sharp. Near the boundary between the stable and unstable regimes in the $\dot{M}_*-P$ plane, accretion through both tongues and funnels is seen. This limits the accuracy of the position of the boundary between the stable and unstable regions.

Decreasing the star's magnetic field increases the extent of the unstable region in \fig{stab-nonstab}. This indicates that smaller accretion rates through the disk are sufficient to trigger the instability. The physical effect of changing the dimensionless magnetic moment $\mu$ is to change the size of the magnetosphere. For $\mu=2$, the ratio of the magnetospheric radius to the stellar radius is between 4 and 5, and for $\mu=0.5$, it is between 2 and 3.

\fignar{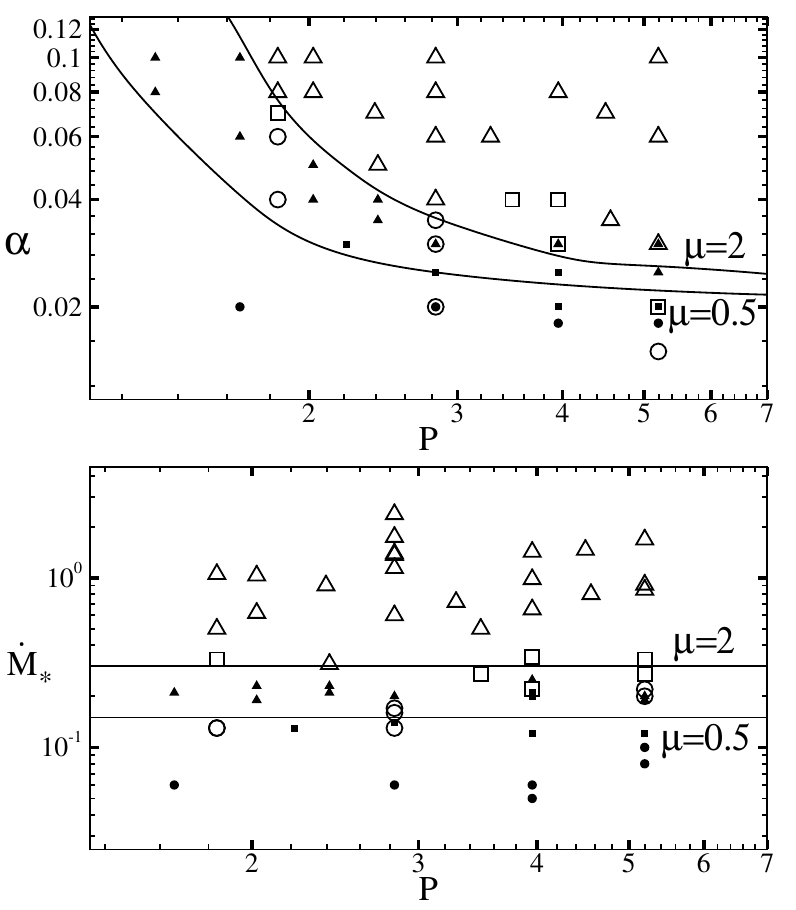}{stab-nonstab}{Regimes of instability in the $\alpha-P$ and $\dot M_*-P$ planes. Two values of the magnetic dipole moment are shown. Triangles, circles and squares represent unstable, stable and borderline cases respectively. The large (small) symbols are for the larger (smaller) dipole moment. The solid lines separate the unstable and stable regions, which are above and below the lines respectively.}

\fig{stab-nonstab} is in dimensionless units, and therefore, as noted in \S\ref{sec:refval}, can be used for a wide variety of physical situations with appropriately chosen reference values. As an illustration, we show here the critical value of the surface accretion rate separating the stable and unstable regimes for $\mu=2$, and the reference values of the star's rotation period, for various physical systems. One must be careful when using the reference values in Table \ref{tab:refval}, because as noted in Appendix \ref{app:refval}, a dimensionless magnetic moment of $\mu=2$ corresponds to a surface magnetic field of $B_*=2B_{*0}$, where $B_{*0}$ is the reference value of the surface magnetic field. From the bottom panel of \fig{stab-nonstab}, the dividing line between the stable and unstable regimes for $\mu=2$ is at $\dot M_{crit} = 0.3$ in dimensionless units, which translates into the following values for classical T Tauri stars (CTTSs), white dwarfs and neutron stars, where the star has mass $M$, radius $R$ and surface magnetic field $B$:

\begin{enumerate}

\item CTTSs:
\begin{eqnarray}
\label{eq-refval-1}
\nonumber
\dot M_{*,crit} &=& 2.1\e{-8}M_\odot \mbox{ yr}^{-1} \times \\
 && \left(\frac{B}{10^3\mbox{ G}} \right)^2 \left(\frac{R}{2R_\odot} \right)^{5/2} \left(\frac{M}{0.8M\odot} \right)^{-1/2} \\
P_0 &=& 1.8\mbox{ days} ~ \left(\frac{R}{2R_\odot} \right)^{3/2} \left(\frac{M}{0.8M\odot} \right)^{-1/2}
\end{eqnarray}

\item White dwarfs:
\begin{eqnarray}
\nonumber
\dot M_{*,crit} &=& 1.4\e{-8}M_\odot \mbox{ yr}^{-1} \times \\
 && \left(\frac{B}{10^6\mbox{ G}} \right)^2 \left(\frac{R}{5000\mbox{ km}} \right)^{5/2} \left(\frac{M}{M\odot} \right)^{-1/2} \\
P_0 &=& 29\mbox{ s} ~ \left(\frac{R}{5000\mbox{ km}} \right)^{3/2} \left(\frac{M}{M\odot} \right)^{-1/2}
\end{eqnarray}

\item Neutron stars:
\begin{eqnarray}
\label{eq-refval-2}
\nonumber
\dot M_{*,crit} &=& 2.2\e{-9}M_\odot \mbox{ yr}^{-1} \times \\
 && \left(\frac{B}{10^9\mbox{ G}} \right)^2 \left(\frac{R}{10\mbox{ km}} \right)^{5/2} \left(\frac{M}{1.4M\odot} \right)^{-1/2} \\
P_0 &=& 2.2\mbox{ ms} ~ \left(\frac{R}{10\mbox{ km}} \right)^{3/2} \left(\frac{M}{1.4M\odot} \right)^{-1/2}
\end{eqnarray}

\end{enumerate}

This is useful because if the mass, radius and magnetic field of the star and the accretion rate are known from observations, one can use equations (\ref{eq-refval-1}) to (\ref{eq-refval-2}) to find out if the star is expected to be in the stable or unstable regime of accretion, a fact which is important for variability and spectral modelling. Conversely, if the nature of the accretion flow (stable or unstable) is well constrained by observations, then equations (\ref{eq-refval-1}) to (\ref{eq-refval-2}) can be used to constrain the star's magnetic field.

The above results have been obtained from simulations of relativistic stars (using the Paczy\'nski-Wiita potential, as mentioned in \S \ref{sec:model}), for the fiducial neutron-star parameters shown in Table \ref{tab:refval}, which correspond to $r_g/R_*=0.4$. However, simulations for non-relativistic stars have shown that the unstable accretion has similar features, and a difference of about 20\% in quantities such as $\dot M$.

\section{Comparison with analytical stability criteria}
In the simple case of a high-density fluid supported against gravity by a low-density fluid with a homogeneous magnetic field, with a plane boundary between them, the development of the Rayleigh-Taylor and Kelvin-Helmholtz instabilities in the direction perpendicular to the field is not affected by the field --- all perturbation modes are unstable (Chandrasekhar 1961). This would suggest that for a star with a dipole field, azimuthal perturbations at the inner disk boundary should always be unstable. However, the inner disk usually has a relatively strong azimuthal field component $B_\phi$, which develops due to the difference between the rotation rate of the star and the Keplerian rotation speed at the inner edge of the accretion disk. An azimuthal field is expected to suppress short wavelength perturbations, and this has been observed in earlier simulations (Wang \& Robertson 1984, 1985; Rast\"atter \& Schindler 1999b; Stone \& Gardiner 2007a,b). The azimuthal field is usually found to be about (5--30)\% as strong as the vertical magnetic field in our simulations. Keeping this in mind, broad conclusions about the instabilities can be drawn from the analytical results for simple cases like the one mentioned above. The Kelvin-Helmholtz instability in the azimuthal direction is suppressed if
\begin{equation}
\label{eq:chandra-kh}
\rho_m(v_d-v_m)^2 \leq \frac{B_\phi^2}{2\pi},
\end{equation}
(Chandrasekhar 1961, \S106), where $\rho$ and $v$ are the plasma density and azimuthal velocity, and the subscripts $d$ and $m$ refer to the disk and the magnetosphere, and we have used $\rho_d>>\rho_m$. This inequality is applicable only for a plane boundary between the fluids, but is sufficient for the rough estimates we are performing here. Rough values of the above quantities from some of our typical simulations, in dimensionless units, are as follows: $\rho_m \sim 0.01$, $v_d-v_m \sim 0.1$ and $B_\phi\approx 0.3$. We thus see that the inequality (\ref{eq:chandra-kh}) is easily satisfied. The Kelvin-Helmholtz instability is completely suppressed.

\fignar{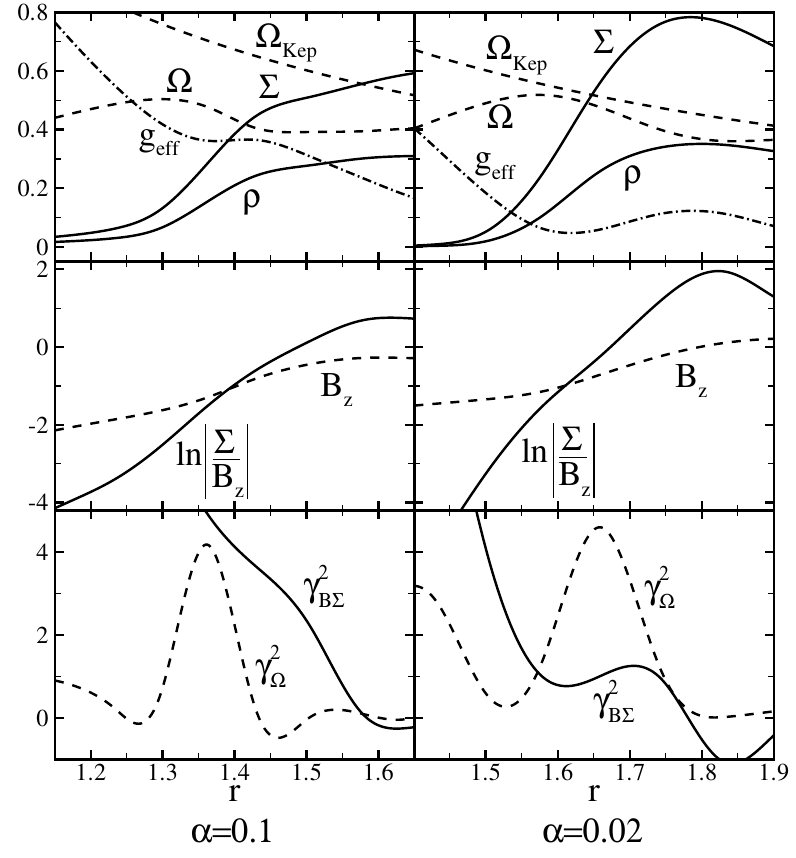}{spruit-alpha}{Various terms in the criterion (\ref{eq:spruit-taam}) for our main case, with $\alpha=0.1$ (left column, unstable) and $\alpha=0.02$ (right column, stable).}

Azimuthal perturbations with wavelength $\lambda$ are Rayleigh-Taylor unstable if
\begin{equation}
\lambda > \frac{B_\phi^2}{\rho_d g_{eff}},
\end{equation}
(Chandrasekhar 1961, \S97), where $g_{eff} \equiv g - \Omega^2r$ is the effective gravitational acceleration ($g$ and $g_{eff}$ are positive if the acceleration is radially inwards). In the linear regime, the amplitudes of the azimuthal perturbation modes $m$ are proportional to $e^{im\phi}$. Then we have $\lambda=2\pi r_m/m$, where $r_m$ is the magnetospheric radius. Thus, unstable modes are those that satisfy
\begin{equation}
m < \frac{2\pi r_m\rho_d g_{eff}}{B_\phi^2}.
\end{equation}
From our simulations, we have, roughly, $r_m \approx 1.5$, $\rho_d \approx 0.7$, $g_{eff} \approx 0.1$, $B_\phi \approx 0.3$, which gives $m \lesssim 7$, for both stable and unstable cases. Thus, this criterion successfully predicts the suppression of high-$m$ Rayleigh-Taylor modes, but not the complete suppression of the instability in some cases. Clearly some other mechanism not accounted for here is responsible for suppressing the instability. One important factor missing from the above analysis is the effect of the radial shear of the angular velocity, which can suppress the instability by smearing out the perturbations. Spruit, Stehle \& Papaloizou (1995) have performed a more general analysis of disk stability in the thin disk approximation, taking the velocity shear into account. The disk has a surface density $\Sigma$ and is threaded by a magnetic field with a vertical component $B_z$. Their criterion for the existence of the instability is
\begin{equation}
\label{eq:spruit-taam}
\gamma_{B\Sigma}^2 \equiv g_{eff} \dd{r} \ln \left| \frac{\Sigma}{B_z} \right| > 2 \left( r \dd[\Omega]{r} \right)^2 \equiv \gamma_\Omega^2.
\end{equation}
In other words, $\Sigma/B_z$ should increase with $r$ fast enough to overcome the stabilizing effect of the velocity shear. \fig{spruit-alpha} shows the relevant quantities for this criterion, azimuthally averaged in the equatorial plane of the disk. The left column shows our main case, which has $\alpha=0.1$ and is unstable. The case in the right column differs from the main case only in that it has $\alpha=0.02$ and is stable. The radial extent shown spans the inner disk region. Note that for $\alpha=0.1$, since the accretion rate through the disk is higher than for $\alpha=0.02$, the inner disk is closer to the star. Due to this, the most significant difference between the top two panels is that the departure of the plasma orbital frequency from Keplerian in the inner disk is larger for the $\alpha=0.1$ case. The other terms in the criterion (\ref{eq:spruit-taam}) do not differ significantly between these two cases. As a result, $\gamma_{B\Sigma}^2$ is much larger for $\alpha=0.1$ than for $\alpha=0.02$, as seen in the bottom two panels of the figure. If one defines the transition region between the disk and the magnetosphere as that over which the plasma density changes from $\rho_d$ to $\rho_m$, then the bottom two panels show that for $\alpha=0.1$, $\gamma_{B\Sigma}^2 > \gamma_\Omega^2$ over almost the entire transition region, predicting instability, while for $\alpha=0.02$, $\gamma_{B\Sigma}^2 < \gamma_\Omega^2$ over most of the transition region, predicting stability. The main driver for the instability in the $\alpha=0.1$ case is thus seen to be the stronger effective gravitational acceleration. However, it is important to note the role that the shear term $\gamma_\Omega^2$ plays here: although it is approximately the same in both cases, it is larger than $\gamma_{B\Sigma}^2$ for $\alpha=0.02$, which suppresses the instability.

We performed a similar comparison for stars with different rotation periods $P$. The right column of \fig{spruit-rc} shows the same case as in the right column of \fig{spruit-alpha}, which has $P=3$ and is stable, and the case in the left column of \fig{spruit-rc} differs from that in the right column only in that it has $P=11$ and is unstable. This time, since $\alpha$ is the same in both cases, the accretion rate through the disk is also the same, and the inner disk region is similar in both cases.
The angular velocity $\Omega$ is almost constant in most of the transition region in both cases. But in the $P=11$ case, in the inner part of the transition region, $\Omega$ drops off sharply towards the star as magnetic braking tries to force the plasma to corotate with the star. This increases the shear term $\gamma_\Omega^2$ in this region. But precisely because the plasma is slowed down, the effective gravity in the inner transition region is higher, with the result that $\gamma_{B\Sigma}^2 > \gamma_\Omega^2$ over almost the entire transition region, once again predicting instability. For the $P=3$ case, once again, the shear is almost the same as in the outer transition region for $P=11$, but the weaker effective gravity reduces $\gamma_{B\Sigma}^2$ to below $\gamma_\Omega^2$ over almost the entire transition region, predicting stability. Thus, criterion (\ref{eq:spruit-taam}) works very well for these cases.

\fignar{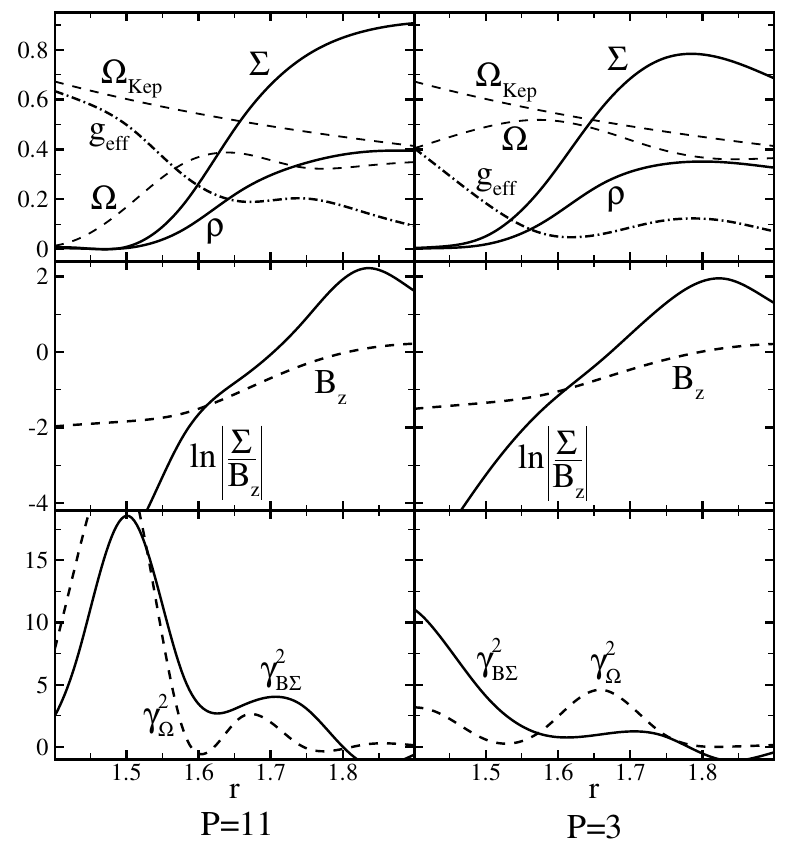}{spruit-rc}{Various terms in the criterion (\ref{eq:spruit-taam}) for our main case, with stellar rotation period $P=11$ (left column, unstable) and $P=3$ (right column, stable).}

Li \& Narayan (2004) have performed linear perturbation analysis on a cylindrically symmetric initial configuration. This analysis is expected to be applicable to the midplane of the disk. They derive analytical stability criteria for this model in a few special cases. One of those cases is where the disk and magnetosphere have constant but different densities, and the angular velocity is constant in the magnetosphere and has a jump at the magnetospheric radius $r_m$. This is the case that is most closely applicable to accretion flows around magnetized stars. The angular velocity profile in the disk is assumed to be $\Omega=\Omega_d(r/r_m)^{-q}$, with $q=0$ or 2. Their criterion for instability is
\begin{eqnarray}
\nonumber
2\left[(1+\mu)\left(m\Omega_{eff,d}^2-\zeta_d^2\right) - (1-\mu)\left(m\Omega_{eff,m}^2+\zeta_m^2\right)\right] \\
\label{eq:li-narayan}
+ (1-\mu^2)[m(\Omega_d - \Omega_m) + \zeta_d + \zeta_m]^2 > 0
\end{eqnarray}
Here,
\begin{eqnarray}
\mu &=& \frac{\rho_d-\rho_m}{\rho_d+\rho_m}, \\
\Omega_{eff}^2 &=& g_{eff}/r,
\end{eqnarray}
and the vorticity,
\begin{equation}
\zeta = \oneby{2r}\dd{r}(r^2\Omega).
\end{equation}
The term on the first line in criterion (\ref{eq:li-narayan}) is the Rayleigh-Taylor term, and the one on the second line is the Kelvin-Helmholtz term. Since $\rho_d>>\rho_m$, $\mu \approx 1$, and (\ref{eq:li-narayan}) becomes
\begin{equation}
m\Omega_{eff,d}^2-\zeta_d^2 > 0.
\end{equation}
The Kelvin-Helmholtz term is again seen to drop out. For unstable modes, we then have,
\begin{equation}
\label{eq:li-narayan-special}
m > \frac{\zeta_d^2}{\Omega_{eff,d}^2} \equiv m_{crit}.
\end{equation}
We see that low vorticity and a high effective gravitational acceleration (measured by $\Omega_{eff}$) are conducive to the development of the instability. \fig{linarayan-alpha} shows the relevant quantities for this criterion in the inner-disk region for the same two simulation runs as in \fig{spruit-alpha}. The quantities are again in the equatorial plane, and azimuthally averaged. As noted earlier, the effective gravity is stronger for $\alpha=0.1$. Also, the vorticity is seen to be slightly smaller. Although this criterion does not work as well as criterion (\ref{eq:spruit-taam}), $m_{crit}$ has a definite tendency to be smaller for $\alpha=0.1$.

\fignar{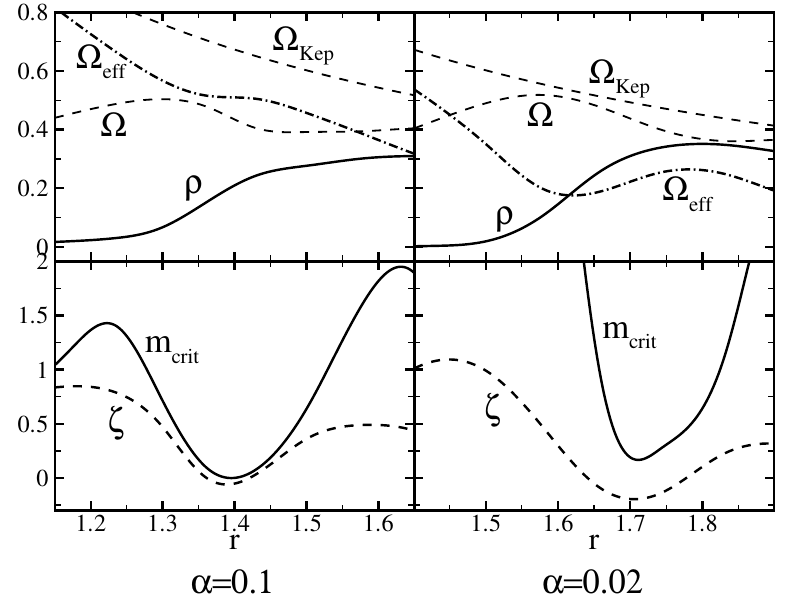}{linarayan-alpha}{Relevant quantities for the criterion (\ref{eq:li-narayan-special}) for our main case, with $\alpha=0.1$ (left column, unstable) and $\alpha=0.02$ (right column, stable).}

\section{Results and Discussion}
Accretion through instabilities at the disk-magnetosphere interface is found to occur for a wide range of physical parameters. It results in tall, thin tongues of gas penetrating the magnetosphere and travelling in the equatorial plane. The tongues are very transient, and grow and rotate around the star on the inner-disk dynamical timescale. The number of tongues at a given time is of the order of a few, irrespective of the grid resolution. Near the star, the tongues are threaded by the magnetic field lines, and form miniature funnel-like flows, which gives them a characteristic wishbone shape. They deposit matter much closer to the star's equator than true funnel flows do. Each tongue produces its own hot spots on the star's surface, and as a result, the hot spots also change on the inner-disk dynamical timescale. Our simulations were performed for stars with magnetospheric radius 2--5 times the radius of the star.

The instability is associated with high accretion rates, and coexists with funnel flows for a relatively broad range of accretion rates. The instability is expected to occur in most accreting systems for typical values of mass, radius, surface magnetic fields and accretion rates.

The instability is suppressed if the misalignment angle $\Theta$ between the star's rotation and magnetic axes is large ($\Theta \gtrsim 30^\circ$). For $\Theta \lesssim 30^\circ$, when the accretion is through instability, the rotational latitude at which most of the accreting matter falls on the star is independent of the misalignment angle.

The number of tongues we see is of the order of a few. The natural question that arises is whether this result is robust with respect to the grid resolution. From theory, we expect small-wavelength Rayleigh-Taylor modes to grow faster, at least in the linear regime (e.g. Chandrasekhar 1961). Solar prominences, which are also subject to the Rayleigh-Taylor instability (Anzer 1969), have a high degree of filamentary structure. If the inner edge of the disk has a similar structure, then increasing the grid resolution would resolve ever thinner tongues, increasing their number. However, we think that this is not the case. Wang \& Robertson (1985) investigated the late stages of the Rayleigh-Taylor instability with a plane boundary between the two fluids, in two dimensions. One of their important observations was that although short wavelength modes initially grow faster, their growth is limited by viscous damping. Further evolution is possible only towards larger spatial scales, which is accomplished by the clumping together of the smaller wavelength structures. We see the same behaviour in our simulations. Two other important stabilizing factors for the Rayleigh-Taylor instability are the tension of the azimuthal magnetic field and the radial shear of the angular velocity. The azimuthal magnetic field develops because the stellar magnetic field threads the disk, and is twisted by the inner-disk plasma, which usually rotates faster than the star. The magnetic field tension is expected to suppress short-wavelength perturbations (Chandrasekhar 1961), and this has been observed in earlier simulations (Wang \& Robertson 1984, 1985; Rast\"atter \& Schindler 1999b; Stone \& Gardiner 2007a,b). Our results confirm these observations. The shear of the angular velocity also has a stabilizing effect, since it tends to smear out the perturbations. These, we think, are the reasons why we do not see a large number of tongues. As a final note, the number of grid cells spanning the width of a fully grown tongue in our highest resolution runs is of the order of 10, which indicates that the grid is fine enough to resolve much thinner tongues than those we see.

Although the shear tends to suppress the Rayleigh-Taylor instability, if the shear is large enough, one would expect the Kelvin-Helmholtz instability to develop. However, the angular velocity profile in the inner disk is generally close to being flat, since the stellar magnetic field tries to force the inner-disk plasma closer to corotation. Rough comparisons with the Kelvin-Helmholtz criterion for a plane boundary between two fluids (Chandrasekhar 1961) suggest that the shear that we observe is not strong enough to overcome the stabilizing effect of the azimuthal magnetic field, although it is strong enough to suppress the Rayleigh-Taylor instability in some cases.

Analytical stability criteria have been derived by a number of authors. We compared our simulations with two of them here. Spruit \& Taam (1995) derived a stability criterion for thin disks, which is found to work reasonably well. Li \& Narayan (2004) derived a stability criterion for a cylindrically symmetric initial equilibrium, which is expected to be applicable in the disk midplane. This criterion works only approximately. This is possibly because the particular angular velocity profiles considered by those authors, and the assumption of a sharp magnetospheric boundary, which are necessary for analytical tractability, do not approximate the conditions in realistic disks as closely as one would wish.

One of the most interesting observational consequences of accretion through instabilities is the effect on the variability. Light curves associated with accretion through funnel streams show clear periodicity, but those associated with accretion through tongues often do not. This of relevance to young stars, whose light curves often lack clear periodicity (e.g. Herbst et al. 2000, 2002). Such light curves would not preclude the existence of an ordered stellar magnetic field if accretion through instabilities were taken into account. On the other hand, we sometimes see that a certain number of tongues dominates, which may lead to quasi-periodic oscillations in the lightcurves (Li \& Narayan 2004). These QPOs may be important for understanding the Type II (accretion-driven) bursts in LMXBs. Comptonization of photons by high-energy electrons may lead to only a small departure of the light-curve from the thermal ones obtained using the approximation of isotropic black-body radiation (Poutanen \& Gierli\'nski 2003) so that the QPO features may survive comptonization (see, however, Titarchuk et al. 2007). From a different angle, the instability found in our simulations may be related to the unstable ``corotation'' mode found by Lovelace \& Romanova (2007) near the disk/magnetosphere boundary by a linear WKB stability analysis.

Due to the stochastic nature of the tongues, it is possible for periods of such quasi-periodic oscillations to alternate with periods of no detectable pulsations. Also, for accretion rates near the critical accretion rate associated with the instability, accretion through both tongues and funnels is observed. This raises the possibility of changes in the accretion rate leading to switching between funnel- and tongue-dominated accretion. These phenomena might be related to the intermittent pulsations observed in some LMXBs (see, e.g., Kaaret et al. 2006; Altamirano et al. 2007; Casella et al. 2007; Galloway et al. 2007; Gavriil et al. 2007). We plan to investigate variability associated with unstable accretion in future work.

The other aspect of variability is spectral variability of CTTSs. The spectral line profiles are dependent upon the density and velocity of the accretion flow (see, e.g., Symington et al. 2005; Kurosawa et al. 2006), and will therefore be affected by instabilities. Recently spectral lines were calculated for CTTSs based on our 3D MHD model of stable accretion through funnel streams using a 3D radiative transfer code (Kurosawa et al. 2007). We plan to undertake similar studies for accretion through instabilities.

In the case of young stars surrounded with gas/dust disks where planets are forming and migrating inward, the tongues may support inward migration, so that the magnetospheric gap, which halts migration (see, e.g., Lin, Bodenheimer \& Richardson 1996; Romanova \& Lovelace 2006; Papaloizou 2007), may form only in the state of stable accretion.

\vspace{.3in}

We thank Drs. A. Koldoba and G. Ustyugova for their contribution to the code development, Drs. R. Lovelace, D. Altamirano, P. Kaaret and C. Thompson for discussions, and the referee for valuable suggestions which improved this paper. NASA provided high-performance computational facilities for this work. The research was partially supported by the NSF grants AST-0507760 and AST-0607135, and the NASA grants NNG05GG77G and NNG05GL49G.

\appendix

\section{Reference Values}
\label{app:refval}

\begin{table}
\begin{tabular}{l@{\extracolsep{0.2em}}l@{}lll}

\hline
&                                                   & CTTSs       & White dwarfs          & Neutron stars           \\
\hline

\multicolumn{2}{l}{$M(M_\odot)$}                    & 0.8         & 1                     & 1.4                     \\
\multicolumn{2}{l}{$R$}                             & $2R_\odot$  & 5000 km               & 10 km                   \\
\multicolumn{2}{l}{$R_0$ (cm)}                      & $4\e{11}$   & $1.4\e9$              & $2.9\e6$                \\
\multicolumn{2}{l}{$v_0$ (cm s$^{-1}$)}             & $1.6\e7$    & $3\e8$                & $8.1\e9$                \\
\multicolumn{2}{l}{$\omega_0$ (s$^{-1}$)}           & $4\e{-5}$   & 0.2                   & $2.8\e3$                \\
\multicolumn{2}{l}{\multirow{2}{*}{$P_0$}}          & $1.5\e5$ s  & \multirow{2}{*}{29 s} & \multirow{2}{*}{2.2 ms} \\
&                                                   & $=1.8$ days &                       &                         \\
\multicolumn{2}{l}{$B_{\star_0}$ (G)}               & $10^3$      & $10^6$                & $10^9$                  \\
\multicolumn{2}{l}{$B_0$ (G)}                       & 43          & $4.3\e4$              & $4.3\e7$                \\
\multicolumn{2}{l}{$\rho_0$ (g cm$^{-3}$)}          & $7\e{-12}$  & $2\e{-8}$             & $2.8\e{-5}$             \\
\multicolumn{2}{l}{$p_0$ (dy cm$^{-2}$)}            & $1.8\e{3}$  & $1.8\e{9}$            & $1.8\e{15}$             \\
\multirow{2}{*}{$\dot M_0$} & (g s$^{-1}$)          & $1.8\e{19}$ & $1.2\e{19}$           & $1.9\e{18}$             \\
                            & ($M_\odot$yr$^{-1}$)  & $2.8\e{-7}$ & $1.9\e{-7}$           & $2.9\e{-8}$             \\
\multicolumn{2}{l}{$T_0$ (K)}                       & $1.6\e6$    & $5.6\e8$              & $3.9\e{11}$             \\
\multicolumn{2}{l}{$\dot E_0$ (erg s$^{-1}$)}       & $4.8\e{33}$ & $1.2\e{36}$           & $1.2\e{38}$             \\
\multicolumn{2}{l}{$T_{\mathrm{eff},0}$ (K)}        & 4800        & $3.2\e5$              & $2.3\e7$                \\
\hline
\end{tabular}
\caption{Sample reference values of the dynamical quantities used in our simulations.}
\label{tab:refval}
\end{table}

As stated in section \ref{sec:refval}, our simulations are done using dimensionless variables, obtained by dividing the dimensional variables by their respective reference values. The reference values are determined as follows: The unit of distance $R_0$ is chosen such that the star has radius $R = 0.35R_0$. The reference velocity is the Keplerian velocity at $R_0$, $v_0 = (GM/R_0)^{1/2}$, and $\omega_0 = v_0/R_0$ is the reference angular velocity.
The reference time is the Keplerian rotation period at $R_0$, $P_0 = 2\pi R_0/v_0$. The reference surface magnetic field of the star at the magnetic equator is $B_{\star_0}$. The reference magnetic field, $B_0$, is the initial magnetic field strength at $r=R_0$, assuming a surface magnetic field of $B_{\star_0}$. The reference magnetic dipole moment is $\mu_0 = B_{*0} R_*^3 \equiv B_0 R_0^3$. A dimensionless magnetic moment of $\mu'$ then corresponds to a surface magnetic field of $B_* = \mu'B_{*0}$. The reference density is taken to be $\rho_0 = B_0^2/v_0^2$. The reference pressure is $p_0 = \rho_0 v_0^2$. The reference temperature is $T_0 = p_0/{\mathcal R}\rho_0$, where $\mathcal{R}$ is the gas constant. The reference accretion rate is $\dot{M}_0 = \rho_0 v_0 R_0^2$. The reference energy flux is $\dot{E}_0 = \rho_0 v_0^3R_0^2$. The reference value for the effective blackbody temperature of the hot spots is $(T_{\mathrm eff})_0 = (\rho_0 v_0^3/\sigma)^{1/4}$, where $\sigma$ is the Stefan-Boltzmann constant. Table \ref{tab:refval} shows sample reference values for three classes of objects: classical T Tauri stars (CTTSs), white dwarfs and neutron stars.


\begin{thebibliography}{}
\bibitem{aetal07} Altamirano, D., Casella, P., Patruno, A., Wijnands, R., van der Klis, M., 2008, ApJ, 674, L45
\bibitem{a69} Anzer, U., 1969, So.Ph., 8, 37
\bibitem{ab80} Anzer, U., B\"orner, G., 1980, A\&A, 83, 133
\bibitem{ab83} Anzer, U., B\"orner, G., 1983, A\&A, 122, 73
\bibitem{al76a} Arons, J., Lea, S. M., 1976a, ApJ, 207, 914
\bibitem{al76b} Arons, J., Lea, S. M., 1976b, ApJ, 210, 792
\bibitem{b77} Baan, W., 1977, ApJ, 214, 245
\bibitem{b79} Baan, W., 1979, ApJ, 227, 987
\bibitem{betal07} Bouvier, J., Alencar, S. H. P., Harries, T. J., Johns-Krull, C. M., Romanova, M. M., 2007, in B. Reipurth, D. Jewitt, K. Keil, eds., Proc. Protostars and Planets V. University of Arizona Press, Tucson, p. 479
\bibitem{cetal07} Casella, P., Altamirano, D., Patruno, A., Wijnands, R., van der Klis, M., 2008, ApJ, 674, L41
\bibitem{c61} Chandrasekhar, S., 1961, Hydrodynamic and Hydromagnetic Stability. Clarendon, Oxford, p. 466
\bibitem{el77} Elsner, R. F., Lamb, F. K., 1977, ApJ, 215, 897
\bibitem{getal07} Galloway, D. K., Morgan, E. H., Krauss, M. I., Kaaret, P., Chakrabarty, D., 2007, ApJ, 654, L73
\bibitem{gvetal07} Gavriil, F. P., Strohmayer, T. E., Swank, J. H., Markwardt, C. B., 2007, ApJ, 669, L29
\bibitem{gl78} Ghosh, P., Lamb, F. K., 1978, ApJ, 223, L83
\bibitem{gl79} Ghosh, P., Lamb, F. K., 1979, ApJ, 232, 259
\bibitem{h98} Hartmann, L., 1998, Accretion Processes in Star Formation. Cambridge Univ. Press, Cambridge
\bibitem{hetal00} Herbst, W., Maley, J. A., Williams, E. C., 2000, AJ, 120, 349
\bibitem{hetal02} Herbst, W., Bailer-Jones, C. A. L., Mundt, R., Meisenheimer, K., Wackermann, R., 2002, A\&A, 396, 513
\bibitem{ketal06} Kaaret, P., Morgan, E. H., Vanderspek, R., Tomsick, J. A., 2006, ApJ, 638, 963
\bibitem{ktl92} Kaisig, M., Tajima, T., Lovelace, R. V. E., 1992, ApJ, 386, 83
\bibitem{ketal02} Koldoba, A. V., Romanova, M. M., Ustyugova, G. V., Lovelace, R. V. E., 2002, ApJ, 576, L53
\bibitem{kr05} Kulkarni, A. K., Romanova, M. M., 2005, ApJ, 633, 349
\bibitem{khs06} Kurosawa, R., Harries, T. J., Symington, N. H., 2006, MNRAS, 370, 580
\bibitem{krh07} Kurosawa, R., Romanova, M. M., Harries, T. J., 2007, submitted to MNRAS
\bibitem{ln04} Li, L.-X., Narayan, R., 2004, ApJ, 601, 414
\bibitem{lbr96} Lin, D. N. C., Bodenheimer, P., Richardson, D. C., 1996, Nature, 380, 606
\bibitem{ls81} Lovelace, R. V. E., Scott, H. A., 1981, in T. D. Guyenne, ed., Proc. International School and Workshop on Plasma Astrophysics (ESA SP-161). ESA, Noordwijk, p. 215
\bibitem{lr07} Lovelace, R. V. E., Romanova, M. M., 2007, ApJ, 670, L13
\bibitem{ls95} Lubow, S. H., Spruit, H. C., 1995, ApJ, 445, 337
\bibitem{nt73} Novikov, I., Thorne, K. S., 1973, in B. DeWitt and C. DeWitt eds., Black Holes. Gordon and Breach, New York, p. 409
\bibitem {p07} Papaloizou, J. C. B., 2007, A\&A, 463, 775
\bibitem{pw80} Paczy\'nski, B., Wiita, P. J., 1980, A\&A, 88, 23
\bibitem{pg03} Poutanen, J., Gierli\'nski, M., 2003, MNRAS, 343, 1301
\bibitem{rn97} Rast\"atter, L., Neukirch, T., 1997, A\&A, 323, 923
\bibitem{rs99a} Rast\"atter, L., Schindler, K., 1999a, ApJ, 519, 658
\bibitem{rs99b} Rast\"atter, L., Schindler, K., 1999b, ApJ, 524, 361
\bibitem{rkl07} Romanova, M. M., Kulkarni, A. K., Lovelace, R. V. E., 2008, ApJ, 673, L171
\bibitem{rl06} Romanova, M. M., Lovelace, R. V. E., 2006, ApJ, 645, L73
\bibitem{retal02} Romanova, M. M., Ustyugova, G. V., Koldoba, A. V., Lovelace, R.V.E., 2002, ApJ, 578, 420
\bibitem{retal03} Romanova, M. M., Ustyugova, G. V., Koldoba, A. V., Wick, J. V., Lovelace, R. V. E., 2003, ApJ, 595, 1009
\bibitem{retal04} Romanova, M. M., Ustyugova, G. V., Koldoba, A. V., Lovelace, R. V. E., 2004, ApJ, 610, 920
\bibitem{ss73} Shakura, N. I., Sunyaev, R. A., 1973, A\&A, 24, 337
\bibitem{st90} Spruit, H. C., Taam, R. E., 1990, A\&A, 229, 475
\bibitem{setal95} Spruit, H. C., Stehle, R., Papaloizou, J. C. B., 1995, MNRAS, 275, 1223
\bibitem{sg07a} Stone, J. M., Gardiner, T. A., 2007a, PhFl, 19, 4104
\bibitem{sg07b} Stone, J. M., Gardiner, T. A., 2007b, ApJ, 2007, 671, 1726
\bibitem{setal05} Symington, N. H., Harries, T. J., Kurosawa, R., 2005, MNRAS, 356, 1489
\bibitem{tks07} Titarchuk, L., Kuznetsov, S., Shaposhnikov, N., 2007, ApJ, 667, 404
\bibitem{vK00} van der Klis, M., 2000, ARA\&A, 38, 717
\bibitem{wn83} Wang, Y.-M., Nepveu, M., 1983, A\&A, 118, 267
\bibitem{wr84} Wang, Y.-M., Robertson, J. A., 1984, A\&A, 139, 93
\bibitem{wr85} Wang, Y.-M., Robertson, J. A., 1985, ApJ, 299, 85
\bibitem{w95} Warner, B., 1995, Cataclysmic variable stars. Cambridge University Press, New York.
\bibitem{ww02} Warner, B., Woudt, P. A., 2002, MNRAS, 335, 84
\end{thebibliography}
\end{document}